\documentclass[11pt]{article}
\usepackage{setspace}
\usepackage{float}
\usepackage{multirow}
\usepackage{slashbox}
\usepackage{natbib}
\bibpunct{(}{)}{;}{a}{}{,}
\usepackage{graphicx}
\usepackage{amssymb}
\usepackage{amsthm}
\usepackage{multirow}
\usepackage{mathrsfs}
\usepackage{geometry}
\usepackage{amsfonts}

\usepackage{url} 
\usepackage{appendix}

\usepackage{geometry}
\usepackage{tikz}
\usetikzlibrary{positioning}
\usetikzlibrary{arrows}
\usetikzlibrary{automata}
\usepackage{syntonly}
\usepackage{tikz}
\usetikzlibrary{positioning}
\usetikzlibrary{arrows}
\usetikzlibrary{automata}

\usepackage[usenames,dvipsnames]{pstricks}
\usepackage{epsfig}
\usepackage{pst-grad}
\usepackage{pst-plot}

\usepackage{epstopdf}
\usepackage{amsmath}
\usepackage{setspace}

\usepackage{lipsum}

\newcommand\blfootnote[1]{%
  \begingroup
  \renewcommand\thefootnote{}\footnote{#1}
  \addtocounter{footnote}{-1}%
  \endgroup
}

\newcommand{\beq}{\begin{equation}}
\newcommand{\eeq}{\end{equation}}

\def\eq#1/{(\ref{e:#1})}
\def\Section#1/{Section~\ref{s:#1}}
\def\Table#1/{Table~\ref{t:#1}}
\def\Figure#1/{Figure~\ref{f:#1}}

\newcommand{\bm}[1]{\mbox{\boldmath$#1$}}

\def\theequation{\arabic{equation}}

\renewcommand{\theequation}{\arabic{equation}}

\input epsf

\begin{document}
\allowdisplaybreaks
\title{\LARGE {\bf A flexible and efficient algorithm for joint imputation of general data}}

\author{
Michael W.~{\sc Robbins}\footnote{Statistician, RAND Corporation, Pittsburgh, PA 15213
(E-mail:~{\em mrobbins@rand.org}).}
\blfootnote{
\textbf{Acknowledgments}:
The author acknowledges funding 
from grant R21AG058123 from the National Institutes of Health.
}
}

\date{August 6, 2020}

\maketitle

\begin{abstract}
Imputation of data with general structures (e.g., data with continuous, binary, unordered categorical, and ordinal variables) is commonly performed with fully conditional specification (FCS) instead of joint modeling.
A key drawback of FCS is that it does not invoke an appropriate data augmentation mechanism and as such convergence of the resulting Markov chain Monte Carlo procedure is not assured. Methods that use joint modeling lack these drawbacks but have not been efficiently implemented in data of general structures. We address these issues by developing a new method, the so-called GERBIL algorithm, that draws imputations from a latent joint multivariate normal model that underpins the generally structured data. This model is constructed using a sequence of flexible conditional linear models that enables the resulting procedure to be efficiently implemented on high dimensional datasets in practice. Simulations show that GERBIL performs well when compared to those that utilize FCS.
Furthermore, the new method is computationally efficient relative to existing FCS procedures.
\\

\noindent {\bf KEY WORDS:} Missing Data, Multiple Imputation, Joint Modeling, Fully Conditional Specification, Chained Equations, Markov Chain Monte Carlo.
\end{abstract}


\section{Introduction} \label{sec1}

Missing data present one of the classical problems of statistical analyses.  Imputation, in which missing values are replaced with plausible entries according to some sort of statistical model, is a highly popular approach for addressing missing data as it yields completed datasets that can be analyzed with traditional techniques.  Modern approaches to imputation have tended to settle within a Bayesian paradigm wherein imputations are sampled at random from a posterior predictive distribution; this begets the multiple imputation framework in which estimators of uncertainty can be adjusted for imputation error through the creation of several imputed datasets.  Most commonly used imputation procedures generate imputations iteratively via Markov chain Monte Carlo (MCMC) in hopes that after a burn-in period of iterations, the imputations will represent draws from the posterior distribution of the missing data given the observed data. Reviews of missing data, imputation, and multiple imputation are numerous---examples include \cite{rubin87a}, \cite{rubin96}, \cite{schafer99}, \cite{carpenter12}, and \cite{little20}.


The current state of the art for missing data problems in large-scale surveys is often considered imputation by fully conditional specification \citep[FCS,][]{raghunathan01, vanbuuren06, vanbuuren10, white11}, also known as chained equations, wherein each variable is imputed from a conditional model that potentially includes all other variables. This process naturally lends itself to imputation of variables of general structure (e.g., binary, unordered categorical, ordinal); further, transformation \citep[e.g.,][]{robbins11, robbins14b, lee17} or predictive mean matching \citep{little88} can be applied to preserve continuous marginal distributions that are non-standard.  Conditional modeling and imputation may be performed with random forests \citep{doove14, shah14} or regression trees \citep{burgette10, doove14} within FCS procedures. However, since the conditional models can be, in theory, incompatible with one another, FCS does not necessarily sample imputations from a valid joint distribution, and as such, 
the imputations are not guaranteed to converge across iterations of MCMC.  In spite of its theoretical flaws, FCS is thought to perform well in practice \citep{lee10, white11, vanbuuren18} and is widely used and available across a host of software \cite[e.g.,][]{raghunathan02, vanbuuren10, su11, honaker11}.

Imputation algorithms that sample from valid joint distributions have been developed \citep[e.g.][]{quartagno12, schafer17, hoff18, zhao18}. However, these procedures tend to be incompatible or highly inefficient with data of general structures or high dimensions. For example, no such procedure includes the flexibility to impose selected conditional dependencies within imputation modeling, which renders such procedures computationally infeasible with data from many large surveys.

Here, we introduce a new procedure that borrows from earlier ideas \citep{carpenter12, robbins13} and addresses the theoretical and empirical issues encountered with FCS.  This new  algorithm, referred to as General Efficient Regression-Based Imputation with Latent processes (GERBIL), imposes a latent multivariate normal process in order to facilitate imputation of continuous, binary, unordered categorical, and ordinal (i.e., ordered categorical) variables.  To ensure theoretical validity, GERBIL draws imputations from a joint model while building that model from a sequence of linear conditional models. Modeling in such a fashion enables flexibility in the selection of conditional relationships that permitted between variables.  The SWEEP operator \citep{goodnight79} optimizes the computational performance of the algorithm.  
GERBIL also has the potential to be dramatically more computationally efficient than FCS with high dimensional data.

\section{An Imputation Primer} \label{sec1a}

We begin by sketching fundamental concepts for imputation of missing data.  Relevant imputation methods are founded on the concept of data augmentation \citep[DA,][]{tanner87}.  DA is designed for cases where the desired objective of sampling from a posterior distribution $P(\theta|y)$ is difficult, but for some latent variable $z$, sampling from $P(z|y,\theta)$ and $P(\theta|y,z)$ is simple, where $P(\cdot)$ is general notation for a probabilistic density.  As such, DA involves iteratively sampling from $P(z|y,\theta)$ and $P(\theta|y,z)$ in order to yield valid draws from $P(\theta,z|y)$. In missing data models, it is common to let $y$ represent the observed data in the DA formulation, $z$ represent the missing data, and $\theta$ model parameters.  As such, imputation via DA involves iteratively alternating between an imputation step (or I Step), which involves sampling updated imputations from the density of the missing data given the observed data and the parameters sampled from the previous iteration, and a parameter step (P Step) wherein one samples parameters from the density of the parameters given the observed data and the imputations sampled from the preceding I Step.

To illustrate the DA process with more formal notation, let $\bm{\chi}_{\rm obs}$ denote the observed data and $\bm{\chi}_{\rm mis}$ denote the missing data, while $\bm{\chi}=\{\bm{\chi}_{\rm obs}, \bm{\chi}_{\rm mis}\}$ gives the complete data.  Furthermore, $\bm{\Theta}$ is a set of model parameters that govern the distribution of $\bm{\chi}$.  The objective is to sample imputations from $P(\bm{\chi}_{\rm mis}|\bm{\chi}_{\rm obs},\bm{\Theta})$. Letting $\bm{\chi}_{\rm mis}^{(t)}$ and $\bm{\Theta}^{(t)}$ represent samples of $\bm{\chi}_{\rm mis}$ and $\bm{\Theta}$ drawn at the $t^{\rm th}$ iteration, these are updated within the $(t+1)^{\rm th}$ iteration as follows: \\

\noindent I Step: Draw $ \bm{\chi}_{\rm mis}^{(t+1)}$ from $P(\bm{\chi}_{\rm mis}|\bm{\chi}_{\rm obs},\bm{\Theta}^{(t)})$. \\
\noindent P Step: Draw $\bm{\Theta}^{(t+1)}$ from $P(\bm{\Theta}|\bm{\chi}_{\rm obs}, \bm{\chi}_{\rm mis}^{(t+1)})$.\\

\noindent As $t \rightarrow \infty$, convergence is observed in that $\{ \bm{\chi}_{\rm mis}^{(t)}, \bm{\Theta}^{(t)}\}$ can be shown to represent a random draw from $P( \bm{\chi}_{\rm mis}, \bm{\Theta}|\bm{\chi}_{\rm obs})$.  Validity of estimators derived from the imputed data is contingent upon the missing at random assumption \citep[in the nomenclature of][]{little20}.

Gibbs sampling \citep{geman84} is used to update imputations within the I Step.  Specifically, letting $\bm{\chi}=\{\bm{X}_1,\ldots,\bm{X}_p\}$, within the $(t+1)^{\rm th}$ iteration, we sequentially update $\bm{X}_j^{(t)}$ for each $j$ by replacing values that were originally missing (in $\bm{X}_j$) with draws from
\begin{equation} \label{gibbs}
P\left( \bm{X}_j \left| \bm{X}^{(t+1)}_1, \ldots, \bm{X}^{(t+1)}_{j-1}, \bm{X}^{(t)}_{j+1}, \ldots, \bm{X}^{(t)}_p, \bm{\Theta}^{(t)} \right. \right),
\end{equation}
which serves to create $\bm{X}_j^{(t+1)}$.  In the event that $\bm{\chi}$ follows a Gaussian distribution, multivariate normal theory can be used to form of each of the above conditional models given a mean vector and covariance matrix extracted from $\bm\Theta^{(t)}$. However, joint modeling in this manner for more general data, which may contain binary, unordered categorical or ordinal variables, is more complicated.  Elaborating, 
one can construct a joint model via a sequence conditional models using
\[
P(\bm{X}_1, \bm{X}_2, \ldots, \bm{X}_p| \bm{\Theta}) = \prod_{j=1}^{p} P(\bm{X}_j|\bm{X}_1, \ldots, \bm{X}_{j-1},\bm{\theta}_j^*),
\]
where $\bm{\Theta}=\{\bm{\theta}_1^*,\ldots,\bm{\theta}_p^*\}$. Given the specific marginal structure of each $\bm{X}_j$, models for
\begin{equation} \label{conditionals}
P(\bm{X}_j|\bm{X}_1, \ldots, \bm{X}_{j-1})
\end{equation}
and $\bm{\theta}_j^*$ may be easily determined for $j = 1,\ldots, p$, which yields a valid joint density.  Nonetheless, sampling from
\begin{equation} \label{conditionals1}
P\left( \bm{X}_j \left| \bm{X}_1, \ldots, \bm{X}_{j-1}, \bm{X}_{j+1}, \ldots, \bm{X}_p \right. \right)
\end{equation} 
for $j = 1, \ldots, p$ in a manner that is congenial with the resulting joint density, as is required for valid Gibbs sampling, 
often presents an intractable problem for general data structures.

Fully conditional specification (FCS) circumvents the above problem by modeling each conditional expression of the form in (\ref{conditionals1}) instead of addressing the joint distribution.  As such, in lieu of a P Step, FCS samples model parameters for each conditional model within each phase of the Gibbs sampling.  That is, for each $j = 1,\ldots,p$, imputations for $\bm{X}_j$ at the $(t+1)^{\rm th}$ iteration are determined via 
%
\begin{eqnarray*}
 \bm{\theta}_j^{(t+1)} &\sim& P(\bm{\theta}_j|\bm{X}^{(t+1)}_1, \ldots, \bm{X}^{(t+1)}_{j-1}, \bm{X}^{(t)}_{j}, \ldots, \bm{X}^{(t)}_p), \\
\bm{X}^{(t+1)}_j &\sim& P(\bm{X}_j|\bm{X}^{(t+1)}_1, \ldots, \bm{X}^{(t+1)}_{j-1}, \bm{X}^{(t)}_{j+1}, \ldots, \bm{X}^{(t)}_p, \bm{\theta}_j^{(t+1)}).
\end{eqnarray*}
\noindent where $\bm{\theta}_j$ indicates model parameters for the density seen in (\ref{conditionals1}).  
Since the sequence conditional expressions given by (\ref{conditionals1}) may define an incoherent joint distribution when modeled separately, there is no guarantee that $\{\bm{\chi}_{\rm mis}^{(t)}, \bm{\Theta}^{(t)}\}$ will converge to $P( \bm{\chi}_{\rm mis}, \bm{\Theta}|\bm{\chi}_{\rm obs})$ across iterations with FCS; in fact, divergence is possible. Most references that discuss convergence in FCS methods \citep[e.g.,][]{white11, vanbuuren18} recommend the use of a small number of iterations of MCMC (usually as low as five, which is the default in several algorithms), perhaps to hedge against the possibility of divergence.

Researchers have noted performance issues with FCS when applied in high dimensional datasets \citep[e.g.,][]{loh16}; nonetheless, it has observed prevalent usage when applied in a large-scale surveys \citep[e.g.,][]{schenker06}.

\section{A Joint Imputation Algorithm for General Data Structures} \label{sec2}

Here, we introduce a novel imputation method: the so-called General Efficient Regression-Based Imputation with Latent processes (GERBIL). This procedure is designed to accomplish the following:
\begin{enumerate}
\item Sample imputations from a coherent joint distribution;
\item Have the flexibility to impute variables of a variety of structures (e.g., continuous, binary, unordered categorical, ordinal);
\item Afford the user the ability to determine which conditional relationships are permitted within the imputation model;
\item Be computationally feasible and efficient in high dimensional datasets.
\end{enumerate}
In this endeavor, we revisit the data augmentation framework, but instead of assuming that the latent process $z$ (as described at the beginning of Section \ref{sec1a}) represents only the missing data whereas the other process $y$ is the observed data, we assume that there is a latent data system that underpins all data values (observed or missing) and that the collected data instead represent available knowledge regarding this system in that some variables may be fully or partially observed.

\subsection{Defining the Latent Process} \label{sec21}

As in Section \ref{sec1a}, let $\bm{\chi}=\{\bm{X}_1,\ldots,\bm{X}_p\}$ denote that collected data (which may contain missing values).  We assume that each variable contained in $\bm{\chi}$ has either a continuous, categorical, binary, or ordinal distribution.  Extensions involving semi-continuous data and right-censored data are discussed in Section \ref{sec5}.  For simplicity, we assume that binary variables take on value 0 or 1, and we assume that if $\bm{X}_j$ is unordered categorical or ordinal with $k_j>2$ possible values, then $\bm{X}_j \in \{1,\ldots, k_j\}$.  We reformat the data so that if $\bm{X}_j$ is unordered categorical, 
it is represented by $k_j-1$ nested binary variables. 
However, missingness is imposed in a nested binary variable for cases where the categorical variable was observed to fall into a category antecedent to the one corresponding to the that binary variable.
To elaborate, a categorical variable $\bm{X}_j$ is reformatted into variables $\bm{X}_{j'}^*,\ldots,\bm{X}_{j'+k_j-2}^*$ for some index $j'$ as follows:  \\
%
 \begin{equation} \label{cate}
\bm{X}_{j'+\ell-1}^* =
\left\{
\begin{array}{ll}
?, & \mbox{if $\bm{X}_j<\ell$ or $\bm{X}_j= \ ?$}, \\
1, & \mbox{if $\bm{X}_j=\ell$}, \\
0, & \mbox{if $\bm{X}_j>\ell$}.
\end{array}\right.
\end{equation}
for $1\leq \ell \leq k_j-1$ where ``?'' indicates a missing value.  All ?s in $\bm{X}_{j'}^*,\ldots,\bm{X}_{j'+k_j-2}^*$ are imputed.
Let $\bm{\chi}^*=\{\bm{X}_1^*,\ldots,\bm{X}_q^*\}$ denote the reformatted data, where $q \geq p$ and where $\bm{\chi}^*$ contains only continuous, binary, and ordinal variables.  Note that variables that are not unordered categorical are copied over from $\bm{\chi}$ to $\bm{\chi}^*$.
For an unordered categorical variable $\bm{X}_j$, we suggest ordering the categories from least to most prevalent when creating the nested variables; this will minimize the number of missing values that are artificially imposed.

The formulation in (\ref{cate}) represents a nested version of the manner in which semi-continuous (i.e., mixed discrete/continuous) data are frequently handled in imputation algorithms \cite[e.g.,][]{robbins13}.  Specifically, the categorical variable is first broken down into two variables: 1) a binary variable that indicates whether or not the original variable falls into the first category and 2) a categorical variable that is set as the value of the original variable but is missing when the original variable falls into the first category.  Next, this second (categorical) variable is dissected in a similar manner---this yields a second binary variable that is unity when the original variable fell into the second category, missing when it fell into the first, and zero otherwise, along with a third variable that is unordered categorical and contains missing values for cases where the original categorical variable fell into one of the first two categories.  This process is repeated until all categories are embodied by nested binary variables. The advantage of this process is that it allows the nested variables to be (conditionally) independent of one another and is easily reversed following imputation.


Borrowing from the idea of probit modeling, akin to how it has been previously applied in imputation settings \citep{carpenter12}, we assume that a multivariate Gaussian distribution underpins $\bm{\chi}^*$.  Specifically, $\bm{\psi}=\{\bm{Z}_1,\ldots,\bm{Z}_q\}$ indicates the underlying latent process.
We assume that $\bm{\psi} \sim N_q(\bm{\mu},\bm{\Sigma})$ for a mean vector $\bm{\mu}$ and variance matrix $\bm{\Sigma}$.  The process of observed data $\bm\chi^*$ is generated from the latent process $\bm\psi$ as follows:

If $\bm{X}_j^*$ is continuous,
\begin{equation} \label{trans}
\bm{X}_j^* =F_j^{-1}(\Phi(\bm{Z}_j))
\end{equation}
where $F_j(\cdot)$ is the marginal cumulative distribution function (CDF) of $\bm{X}_j^*$, in that $F_j(x)=\mbox{Pr}(\bm{X}_j \leq x)$ where $\mbox{Pr}(A)$ gives the probability of event $A$, and where $\Phi(\cdot)$ denotes the CDF of a standard normal random variable. Of course, prior to imputation, the observed data should be transformed to have a standard normal distribution via the inverse transformation $\bm{Z}_j = \Phi^{-1}(F_j(\bm{X}_j^*))$.   Transformations of this type may be performed with a parametric density \citep{robbins11, robbins13} or in a non-parametric manner with a kernel or empirical distribution \citep{robbins14b}.  This formulation serves to link the continuous data via a Gaussian copula \citep{nelsen09}.
%

If $\bm{X}_j^*$ is binary, a probit-type model is imposed:
\[
\bm{X}_j^* =
\left\{
\begin{array}{ll}
0, & \mbox{if $\bm{Z}_j < 0$}, \\
1, & \mbox{if $\bm{Z}_j \geq 0$}.
\end{array}\right.
\]
%
Lastly, if $\bm{X}_j^*$ is ordinal 
where $\bm{X}_j^* \in \{1,2,\ldots,k_j\}$,
\[
\bm{X}_j^* = i ~~~ \mbox{if $\tau_{j,i-1} < \bm{Z}_j \leq \tau_{j,i}$,}
\]
for $i\in\{1,\ldots,k_j\}$,
where $\tau_{j,i} = \Phi^{-1}(\mbox{Pr}(\bm{X}_j^* \leq i))$ for $i \in \{1,\ldots,k_j-1\}$ and where we set $\tau_{j,0}=-\infty$ and $\tau_{j,k_j}=\infty$.  

Note that the latent multivariate normal process can be modeled conditionally upon a set of fully observed predictors; 
these variables can obey any distribution and need not be underpinned by a Gaussian density.  For simplicity of the exposition, we do not condition on such variables in the following.

\subsection{Imputation of the Latent Process} \label{sec32}

\underline{\em The P Step} of GERBIL builds upon ideas presented in \cite{robbins13}, which addressed missingness in continuous variables.  The objective of the P Step is to determine values of the mean vector $\bm{\mu}$ and variance matrix $\bm{\Sigma}$ of the latent multivariate Gaussian process; however, these quantities are modeled indirectly.  Specifically, we build a joint model for $\bm{\psi}$ by stating linear forms for conditional models seen in (\ref{conditionals}) in that $\bm{Z}_j$ is allowed to depend on variables that precede it in sequence but not those that antecede it.  That is, we assume
\begin{equation} \label{SRmodel}
\bm{Z}_j =  \bm{V}_j \bm{\beta}_j + \sigma_j \bm{\epsilon}_j,
\end{equation}
for $j=1,\ldots,q$, where $\bm{V}_j$ denotes an $n \times \kappa_j$ predictor matrix of which the columns are some subset of the columns of $\{\bm{1}; \bm{Z}_1; \ldots; \bm{Z}_{j-1}\}$, with $\bm{1}$ indicating a vector of ones, and where $\bm{\beta}_j$ denotes a length-$\kappa_j$ vector of regression coefficients---the flexibility to selectively reduce the size of the predictor set for each conditional model is crucial in our setting as referenced previously.  This model imposes that $P(\bm{Z}_j|\bm{Z}_1, \ldots, \bm{Z}_{j-1}) = P(\bm{Z}_j|\bm{V}_{j})$. Note that the predictor matrix $\bm{V}_j$ for a $\bm{Z}_j$ that corresponds to a nested binary variable within an unordered categorical variable should exclude any other nested variables from that same categorical variable.  

Assuming a non-informative prior for $\bm{\Theta}= \{ \bm{\beta}_1, \sigma_1, \ldots, \bm{\beta}_q, \sigma_q \}$ in that $P(\bm{\Theta})\propto \prod^q_{j=1}1/\sigma_j^2$, the posterior distributions of $\bm{\beta}_j$ and $\sigma_j^2$ (assuming fully observed $\bm\psi$) are derived as follows.
If $\bm{X}_j^*$ is binary,
we fix $\sigma_j^2=1$, which is in accordance with traditional probit modeling.  Otherwise,
\begin{equation} \label{post1}
\sigma_j^2 | \bm{\psi} \sim \mbox{Inv-}\chi^2(n-\kappa_j,s_j^2),
\end{equation}
where, letting the superscript $T$ indicate a matrix transpose, $s_j^2 = (\bm{Z}_j-\bm{V}_j\bm{\hat\beta}_j)^{T}(\bm{Z}_j-\bm{V}_j\bm{\hat\beta}_j)/(n-\kappa_j)$ with $\bm{\hat\beta}_j = (\bm{V}_j^{T}\bm{V}_j)^{-1}\bm{V}_j^{T}\bm{Z}_j$ and with $\mbox{Inv-}\chi^2(\:\cdot\:,\:\cdot\:)$ denoting an inverse chi-square distribution.  Likewise,
\begin{equation} \label{post2}
\bm{\beta}_j | \sigma_j^2,\bm{\psi} \sim \mbox{N}_{\kappa_j}(\bm{\hat\beta}_j,\sigma_j^2(\bm{V}_j^{T}\bm{V}_j)^{-1}).
\end{equation}
Given imputed values of the latent process, $\bm{\psi}^{(t)}=\{\bm{Z}_1^{(t)};\ldots;\bm{Z}_q^{(t)}\}$ at the $t^{\rm th}$ iteration, the P Step involves sampling $\bm{\beta}_j^{(t)}$ and $\sigma_j^{(t)}$ from $P(\bm\beta_j,\sigma_j|\bm{Z}_1^{(t)},\ldots, \bm{Z}_{j-1}^{(t)})$ for $j = 1,\ldots,q$ in accordance with (\ref{post1}), when needed, and (\ref{post2}) above.
%

We next calculate $\bm{\mu}^{(t)}$ and $\bm{\Sigma}^{(t)}$, the mean vector and covariance matrix of the process $\bm{\psi}$ at the $t^{\rm th}$ iteration, from the parameter set $\{ \bm{\beta}^{(t)}_1, \sigma_1^{(t)}, \ldots, \bm{\beta}_q^{(t)}, \sigma_q^{(t)} \}$; Section \ref{dets1} of the Supplemental Materials provides illustration of such calculations.  
\\

\noindent
\underline{\em The I Step} for the $(t+1)^{\rm th}$ of GERBIL involves sampling $\bm{\psi}^{(t+1)}$ from $P(\bm{\psi}|\bm{\chi}^*_{\rm obs},\bm{\mu}^{(t)}, \bm{\Sigma}^{(t)})$, where $\bm{\chi}^*_{\rm obs}$ includes the fully and partial observed information regarding $\bm\psi$ from $\bm\chi^*$.
Since $\bm\chi^*$ is uniquely determined from $\bm\psi$, we do not need to recalculate $\bm\chi^*$ at each iteration in order to align with the data augmentation framework.
First, we use $\bm{\mu}^{(t)}$ and $\bm{\Sigma}^{(t)}$ to find the parameters that define
$
P(\bm{Z}_j|\bm{Z}_1,\ldots,\bm{Z}_{j-1},\bm{Z}_{j+1},\ldots,\bm{Z}_p)
$
for each $j=1,\ldots,q$, which is Gaussian since $\bm{\psi}$ multivariate normal.  We execute Gibbs sampling from this distribution.  For each $j \in \{1,\ldots,q\}$,  let
\begin{eqnarray}
\nonumber \mu_{j|\cdot}^{(t+1)} &=& E[\bm{Z}_j|\bm{Z}_1^{(t+1)},\ldots,\bm{Z}_{j-1}^{(t+1)},\bm{Z}_{j+1}^{(t)},\ldots,\bm{Z}_p^{(t)},\bm{\mu}^{(t)},\bm{\Sigma}^{(t)}], \\
\nonumber \sigma_{j|\cdot}^{(t+1)} &=& \mbox{Var}(\bm{Z}_j|\bm{Z}_1^{(t+1)},\ldots,\bm{Z}_{j-1}^{(t+1)},\bm{Z}_{j+1}^{(t)},\ldots,\bm{Z}_p^{(t)},\bm{\mu}^{(t)},\bm{\Sigma}^{(t)}).
\end{eqnarray}
Multivariate normal theory is used to determine $\mu_{j|\cdot}^{(t+1)}$ and $\sigma_{j|\cdot}^{(t+1)}$. Details are provided in Section \ref{dets1} of the Supplemental Materials. 

If $\bm{X}_j^*$ is continuous:
\begin{itemize}
\item For cases where $\bm{X}_j^*$ is observed, set $\bm{Z}_j^{(t+1)} = \bm{Z}_j$;
\item For cases where $\bm{X}_j^*$ is missing, sample $\bm{Z}_j^{(t+1)}$ from $\mbox{N}(\mu_{j|\cdot}^{(t+1)}, \sigma_{j|\cdot}^{(t+1)})$.
\end{itemize}
Note that if $\bm{X}_j^*$ is binary or ordinal, only partial information is known regarding $\bm{Z}_j$, even for cases where $\bm{X}_j^*$ is observed.  This information is incorporated in the sampling scheme for binary $\bm{X}_j^*$ as follows:
\begin{itemize}
\item For cases where $\bm{X}_j^*$ is missing, sample $\bm{Z}_j^{(t+1)}$ from $\mbox{N}(\mu_{j|\cdot}^{(t+1)}, \sigma_{j|\cdot}^{(t+1)})$;
\item For cases with $\bm{X}_j^* =0$, draw $\bm{Z}_j^{(t+1)}$ from $\mbox{trN}(\mu_{j|\cdot}^{(t+1)}, \sigma_{j|\cdot}^{(t+1)},-\infty,0)$;
\item For cases with $\bm{X}_j^*=1$, draw $\bm{Z}_j^{(t+1)}$ from $\mbox{trN}(\mu_{j|\cdot}^{(t+1)}, \sigma_{j|\cdot}^{(t+1)},0, \infty)$.
\end{itemize}
In the above, $\mbox{trN}(\mu,\sigma^2,a,b)$ is a truncated normal distribution with mean $\mu$, variance $\sigma^2$, and bounds of $a$ and $b$.  That is, $X \sim \mbox{trN}(\mu,\sigma^2,a,b)$ implies $X \equiv (Z|a<Z<b)$ with $Z\sim N(\mu, \sigma^2)$.
To find $\bm{Z}_j^{(t+1)}$ if $\bm{X}_j^*$ is ordinal with $k_j$ categories:
\begin{itemize}
\item For cases where $\bm{X}_j^*$ is missing, sample $\bm{Z}_j^{(t+1)}$ from $\mbox{N}(\mu_{j|\cdot}^{(t+1)}, \sigma_{j|\cdot}^{(t+1)})$;
\item For cases with $\bm{X}_j^*=i$ where $1 \leq i \leq k_j$, draw $\bm{X}_j^{(t+1)}$ from $\mbox{trN}(\mu_{j|\cdot}^{(t+1)}, \sigma_{j|\cdot}^{(t+1)},\tau_{j,i-1},\tau_{j,i})$.
\end{itemize}
Herein, we again set $\tau_{j,0}=-\infty$ and $\tau_{j,k_j}=\infty$.

To initialize the MCMC procedure, we find that setting $\mu_{j|\cdot}^{(0)}=0$ and $ \sigma_{j|\cdot}^{(0)}=1$ and sampling $\bm{\psi}^{(0)}=\{\bm{Z}_1^{(0)};\ldots;\bm{Z}_q^{(0)}\}$ according to the rules above performs sufficiently well.  Of course, more rigorous options could be implemented.

\subsection{Derivation of Final Imputations} \label{sec33}

After a burn-in period of $b$ iterations, the MCMC procedure is stopped and ${\bm\psi}^{(b)}=\{\bm{Z}_1^{(b)},\ldots,\bm{Z}_{q}^{(b)}\}$ indicates the final imputed version of the latent data. The final imputations for the (reformatted) recorded dataset are denoted $\widetilde{\bm{\chi}}^*=\{\widetilde{\bm{X}}_1^*,\ldots,\widetilde{\bm{X}}_q^*\}'$ and are derived from ${\bm\psi}^{(b)}$ as follows.

If $\bm{X}_j^*$ is continuous, $\widetilde{\bm{X}}^*_j =F_j^{-1}(\Phi(\bm{Z}_j^{(b)}))$; see \cite{robbins13} and \cite{robbins14b} for specifics regarding transformation and untransformation of marginal distributions.
If $\bm{X}_j^*$ is binary,
\[
\widetilde{\bm{X}}^*_j =
\left\{
\begin{array}{ll}
0, & \mbox{if $\bm{Z}_j^{(b)} < 0$}, \\
1, & \mbox{if $\bm{Z}_j^{(b)} \geq 0$},
\end{array}\right.
\]
and if $\bm{X}_j^*$ is ordinal with $k_j$ categories,
\[
\widetilde{\bm{X}}^*_j = i ~~~ \mbox{if $\tau_{j,i-1} < \bm{Z}_j^{(b)} \leq \tau_{j,i}$}.
\]
for $i \in \{1,\ldots,k_j\}$.

The nesting structure described in (\ref{cate}), in which case an unordered categorical variable $\bm{X}_j \subseteq \bm{\chi}$ from the original datatset has been represented by $\{\bm{X}_{j'}^*,\ldots,\bm{X}_{j'+k_j-2}^*\}\subseteq \bm{\chi}^*$ for some $j'$ in the expanded dataset, is then reversed. This creates the final imputed dataset $\widetilde{\bm\chi}=\{\widetilde{\bm{X}}_1,\ldots,\widetilde{\bm{X}}_p\}$, which is accomplished after setting
\[
\widetilde{\bm{X}}_j =
\left\{
\begin{array}{ll}
1, & \mbox{if $\bm{X}_{j'}^*=1$}, \\
2, & \mbox{if $\bm{X}_{j'+1}^*=1$ and $\bm{X}_{j'}^*=0$}, \\
\vdots & \\
k_j-1,     & \mbox{if $\bm{X}_{j'+k_j-2}^*=1$ and $\bm{X}_{i}^*=0$ for each $i \in \{j',\ldots,j'+k_j-3\}$}, \\
k_j,     & \mbox{if $\bm{X}_{i}^*=0$ for each $i \in \{j',\ldots,j'+k_j-2\}$},
\end{array}\right.
\]
for all categorical $\bm{X}_j$ and setting other variables contained in $\bm\chi$ equal to their corresponding imputed version in $\widetilde{\bm\chi}^*$.

To apply multiple imputation \citep{rubin87a, rubin96}, the entire process illustrated above is repeated independently $m$ times to procedure $m$ separately imputed datasets. Well known combining rules are used to aggregate the datasets and adjust estimators for imputation error.

Note that the marginal transformations that are applied to continuous variables in (\ref{trans}) assume that $F_j(x)=\mbox{Pr}(\bm{X}_j \leq x)$ is known for each relevant $j$ and likewise that $\tau_{j,i} = \Phi^{-1}\{\mbox{Pr}(\bm{X}_j^* \leq i)\}$ is assumed known for each ordinal $\bm{X}_j$.  In practice, these quantities are estimated which may induce bias into the transformations in missingness mechanisms that are not missing completely at random \citep[borrowing the terminology of][]{little20}. However, earlier studies involving continuous data \citep{robbins13,robbins14b} find no evidence of substantial bias stemming from transformations.
Note also that the copula framework applied to continuous variables requires that following the marginal transformations, the transformed variables obey a multivariate normal distribution (i.e., relationships between variables are linear).  The aforementioned studies \citep[e.g.,][]{robbins13,robbins14b} have also shown that in practice, bivariate relationships are often more linear following such transformations than before.

The manner in which we handle categorical variables is, by our knowledge, novel. Alternative approaches proposed by other authors do not impose missingness in nested variables 
\citep{allison02, honaker11, carpenter12}---imputed values of the categorical variable are then set as the category that observes the highest value among the imputed nested variables. However, rigorous evaluations of this approach are scarce, as noted by \cite{carpenter12}.  In contrast, our proposed approach performs well in simulations (see Section \ref{sec3}).

\subsection{The Sweep Operator} \label{sec34}

The sweep operator \citep{beaton64, goodnight79} is used to dramatically improve the computational efficiency of the GERBIL algorithm in both the P Step and I Step.  Specifically, through the use of this operation in the P Step, all information needed for the conditional models of $\bm{Z}^{(t)}_j$ for each $j = 1,\ldots,q$, as seen in (\ref{SRmodel}), can be calculated through nearly the same amount of computations as would be needed to determine only the quantities necessary for the conditional model for $\bm{Z}^{(t)}_q$.  To elaborate, the sweep operator is a matrix transformation that is applied to a specific column of a symmetric matrix (i.e., ``sweeping in'' the column), and groups of columns may be ``swept in'' by applying the operation to the individual columns (and resulting matrices) in sequence.  The operation functions so that columns may be``swept in'' in any given order without changing the end result.

Assume for the moment that each predictor matrix used for the models in (\ref{SRmodel}) contains the maximum permissible number of predictors (i.e., $\bm{V}_j^{(t)}=\{\bm{1},\bm{Z}_1^{(t)},\ldots,\bm{Z}_{j-1}^{(t)}\}$. 
If the first $j$ columns of the $(q+1)\times(q+1)$ matrix $\bm{A}^{(t)}=(\bm{V}_{q+1}^{(t)})'\bm{V}_{q+1}^{(t)}$ are swept in, the result, which is a $(q+1)\times(q+1)$ matrix that we notate with $\bm{B}^{(t)}_j$, 
contains submatrices which yield the information needed to determine $\hat{\bm{\beta}}_j^{(t)}$ and $(s_j^{(t)})^2$ without further matrix computations.  Then, the sweep operator can be applied to column $j+1$ of $\bm{B}^{(t)}_j$
to yield $\hat{\bm{\beta}}_{j+1}^{(t)}$ and $(s_{j+1}^{(t)})^2$.  As such, applying the operation in sequence to columns 1 through $q$ of the matrix $\bm{A}^{(t)}$ yields $\hat{\bm{\beta}}_{j}^{(t)}$ and $(s_{j}^{(t)})^2$ for each $j = 1,\ldots,q$ in the same number of computations it would take to calculate only $\hat{\bm{\beta}}_{q}^{(t)}$ and $(s_{q}^{(t)})^2$.  Note that there also exists a reverse sweep operator that is used to ``sweep out'' any predictors that have been excluded from specific conditional models.

In the I Step, the reverse sweep operator is applied to each of the columns of $(\bm{\Sigma}^{(t)})^{-1}$ separately to help find $\mu_{j|\cdot}^{(t)}$ and $\sigma_{j|\cdot}^{(t)}$.

%

\subsection{Comparisons to Existing Methods}

GERBIL applies joint modeling which avoids the theoretical issues encountered with FCS and guarantees that GERBIL imputations will converge across iterations of MCMC.  That is, the use of joint modeling gives GERBIL a strong advantage over all implementations of FCS (e.g., \texttt{mice}, \texttt{mi}, IVEware) regardless of the conditional model used for imputation. Further, strategic use of the sweep operator in GERBIL ensures that it may be more computationally efficient than existing FCS software.  In addition, most current implementations of imputation by joint modeling \citep[e.g.][]{robbins13, schafer17, zhao18} do not facilitate general data structures.

The R package \texttt{jomo} \citep{carpenter12, quartagno12}, which uses a latent Gaussian process to underpin non-continuous variables, is perhaps most closely aligned with GERBIL in terms of utility, but GERBIL has a number of operational advantages over \texttt{jomo}.
Specifically, \texttt{jomo} does not build the joint model from a sequence of conditional models as seen in (\ref{SRmodel}) but instead attempts to directly estimate the covariance matrix. However, estimation of a covariance matrix that is subject to restrictions (e.g., the diagonal elements that correspond to binary variables must be set to 1) is difficult in practice as the result may not be positive semi-definite.  \texttt{jomo} addresses this issue by using a guess-and-check Metropolis-Hastings algorithm, and further applies a guess-and-check method in lieu of sampling from a truncated normal distribution.  These issues lead to infeasibility of the algorithm when applied to high dimensional, complex data.  Lastly, \texttt{jomo} does not let its user specify dependencies (which is crucial for many real world datasets)---collinearities in the data may render estimation of the covariance matrix infeasible. 

The GERBIL procedure provides a more natural method by which covariance matrices of the latent process can be estimated.  By setting the conditional error variance of the models for binary variables to be one (instead of attempting to restrict diagonal elements of a covariance matrix to be one), we ensure that the resulting covariance matrix will be positive semi-definite and can be estimated using appropriate Bayesian techniques.  Furthermore, variables can be dropped from specific conditional models in (\ref{SRmodel}) while maintaining a positive semi-definite covariance matrix, enabling the user to avoid multi-collinearities and impose desired conditional dependence structures.

Furthermore, \cite{hoff18} introduces a rank-based approach to estimating parameters of a copula model that underpins general data.  This method may be easily extended in order to impute missing data and is implemented for such purpose in the R package \texttt{sbgcop} \citep{hoff18}. This approach is theoretically similar to that of \texttt{jomo} and as such contains some of the same drawbacks (e.g., lack of flexibility regarding which dependencies are enabled, which may lead to its use being infeasible in high dimensional data from complex surveys). Unlike \texttt{jomo}, however, \texttt{sbgcop} circumvents the need to restrict variances that correspond to binary variables to unity through sampling latent data via a correlation matrix (although covariances are indeed estimated through the Gibbs sampling process). 
Additionally, \texttt{sbgcop} does not directly enable the imputation of unordered categorical variables.

\section{Simulations} \label{sec3}

In this section, we perform a simulation study to evaluate the effectiveness of GERBIL and compare its performance to that of several existing procedures.  
The synthetic data are not designed to favor any particular method(s) but are instead designed to be general and applicable for all methods.

First, we generate a dataset that contains six variables with differing marginal structures, loosely outlined as follows:
\begin{itemize}
\item $\bm{X}_1$ -- Unordered categorical
\item $\bm{X}_2$ -- Continuous (fully observed)
\item $\bm{X}_3$ -- Continuous
\item $\bm{X}_4$ -- Binary (generated from a probit-type model)
\item $\bm{X}_5$ -- Ordinal (generated from a probit-type model)
\item $\bm{X}_6$ -- Binary (generated from a logistic model)
\end{itemize}
Elaborating, $\bm{X}_1$ is generated from a multinomial distribution with 4 categories.  A latent process that underpins $\bm{X}_2 \ldots,\ldots \bm{X}_5$ is generated from a multivariate normal distribution with while conditioning on $\bm{X}_1$.  Further, $\bm{X}_6$ is generated from a logistic model conditional on $\bm{X}_1 \ldots,\ldots \bm{X}_5$.  Non-negligible associations exist between all variables.  We generate $n=2,000$ observations of each variable.

Missingness is stochastically imposed in the synthetic data using the following three mechanisms.  In each case, around a third of the observations are missing (excluding $\bm{X}_2$).
\begin{enumerate}
\item MCAR: Missingness probabilities are independent of any other data characteristics.
\item MAR: Missingness probabilities depend upon only the fully observed variable ${\bm X}_2$.
\item NMAR: Missingness probabilities in variable $\bm{X}_j$ depend upon only $\bm{X}_j$ for $j \in \{1,3,\ldots,6\}$.
\end{enumerate}
These mechanisms are designed in line with the nomenclature of \cite{little20}. Further details on the data generating and missingness mechanisms are provided in the Section \ref{simdets} of the supplemental materials.  Note that missingness rates in each variable (with the exception of $\bm{X}_2$) are approximately 33\% under each mechanism.

Next, the missing values are imputed using six distinct methods, three of which utilize FCS, whereas the others implement joint modeling.  Specifically, comparisons to FCS are performed using the implementation available in the R package \texttt{mice} \citep{vanbuuren10}. 
Within \texttt{mice}, one can assign different methods of imputation to each variable, with Gaussian imputation available for continuous variables, logistic regression for binary variables, and polytomous regression for categorical variables.  \texttt{mice} also implements predictive mean matching \cite[PMM,][]{little88}, which uses a nearest neighbor-type approach based on a predictive model and is often applied to handle continuous variables that may have non-Gaussian marginal distributions, as well as classification tress and random forests.  These techniques can also be applied to binary, unordered categorical, and ordinal variables within \texttt{mice}. We also compare against the R packages \texttt{jomo} and \texttt{sbgcop}, both of which employ joint modeling (as described in Section \ref{existing}). In summary, the various methods used for imputation in the simulations are:
\begin{enumerate}
\item jomo: The \texttt{jomo} package is used for imputation (2 seconds per 100 iterations when applied within this simulation setting).
\item sbgcop: The \texttt{sbgcop} package is used for imputation (1.1 seconds per 100 iterations). The categorical variable is handled in accordance with (\ref{cate}).
\item Logistic: \texttt{mice} is used with logistic regression for binary variables, polytomous regression for categorical variables, ordered logistic regression for ordinal variables, and Gaussian imputation for continuous variables (20 seconds per 100 iterations).
\item PMM: \texttt{mice} is used with PMM for all variables (3 seconds per 100 iterations).
\item CART: \texttt{mice} with imputation by classification trees is performed for all variables (1 minute per 100 iterations).
\item GERBIL: General Efficient Regression-Based Imputation with Latent processes as proposed in Section \ref{sec2}  
    (3.8 seconds per 100 iterations).
\end{enumerate}
We also considered \texttt{mice} with random forests (8 minutes per 100 iterations), but due to its computational burden, it was excluded from the larger simulation study. However, abbreviated simulations show it does not perform as well as the other \texttt{mice} methods used here.  The computing times listed are performed on a Windows machine with a 2.8 GHz processor and 32.0 GB of RAM. Note that, due to its use of the SWEEP operator, GERBIL will improve in computational efficiency in comparison to the \texttt{mice} methods as the dimensionality of the data increases.

We use 15 iterations of MCMC for the \texttt{mice} methods, 60 iterations for GERBIL and \texttt{jomo}, and 120 iterations for \texttt{sbgcop}; more iterations of the non-\texttt{mice} methods are used because of their relative computational efficiency and because \texttt{mice} is shown to converge somewhat quicker in the setting of these simulations.  All possible inter-variable dependencies are enabled for the \texttt{mice} methods and GERBIL.  To adjust for imputation error, we use multiple imputation \citep{rubin87a, rubin96} with $m=40$ independently imputed datasets for each method.  This selection of $m$ is in line with the recommendations of \cite{graham02} given the missingness rates used here.

We use $N=5,000$ replications for this simulation study---that is, the above process of simulating and imputing data is repeated independently 5,000 times. The following parameters are tracked in each replication for each method:
\begin{itemize}
\item Means and the variance-covariance matrix of the dataset $\{{\bm X}_{1,1},\ldots,{\bm X}_{1,4},{\bm X}_2,\ldots,{\bm X}_6\}$ where the ${\bm X}_{1,k}$ for $k\in\{1,\ldots,4\}$ are categorical indicators underpinning ${\bm X}_1$ (although the mean and variance of ${\bm X}_2$ are excluded).  There are 8 mean parameters calculated with 8 variances and 36 covariances.
\item Estimated regression coefficients, and standard errors of those coefficients, for all fully specified conditional models of the form $P({\bm X}_j|{\bm X}_1\ldots,{\bm X}_{j-1},{\bm X}_{j+1},{\bm X}_6)$ for $j\in\{1,\ldots,6\}$.  For continuous and ordinal variables, we fit a basic linear model.  For binary variables, we fit a logistic regression, and for the categorical variable, we fit a multinomial log-linear model via the \texttt{nnet} package in R \citep{venables02}.  There are 58 regression parameters tabulated with 58 standard errors on those parameters.
\end{itemize}
We calculate root-mean squared error (rMSE) for all parameters and coverage rates for a subset of parameters.

We let $\hat\theta^{[r]}(x)$ denote the value of a parameter $\theta$ estimated at the $r^{\rm th}$ replication for imputation method $x$ ($\hat\theta^{[r]}(x)$ is calculated as the average of separate estimates of $\theta$ produced for each of the multiply imputed datasets).
For method $x$, we calculate the rMSE in the estimate of $\theta$ as follows:
\[
\mbox{rMSE}_\theta(x) = \sqrt{\frac{1}{N}\sum^{N}_{r=1}[\hat\theta^{[r]}(x)-\theta]^2}.
\]
The rMSE is calculated for all parameters listed above.

We compare the rMSE of GERBIL to the rMSE of each of the competing methods.  Table \ref{tab0} shows the portion of the 168 parameters for which GERBIL yields the better (i.e., smaller) rMSE for each method under each missingness mechanism. We see that in all cases, GERBIL performs better for a majority of the parameters.  The exact rMSE seen for each of the six methods under all missingness mechanisms is reported in the tables seen in Section \ref{simtabs} of the supplemental materials.

\begin{table}
\caption{\label{tab0}
The portion of the 168 parameters for which the rMSE for the respective method in the respective missingness mechanism is greater than the rMSE yielded by GERBIL.
}
\centering
\vspace{.1in}
\setlength{\tabcolsep}{.4em}
\begin{tabular}{lccccc}
\hline \hline
 & sbgcop & jomo & Logistic & PMM & CART \\  \hline
MCAR & 0.643 & 0.613 & 0.595 & 0.601 & 0.738 \\
MAR & 0.690 & 0.661 & 0.589 & 0.655 & 0.762 \\
NMAR & 0.589 & 0.595 & 0.619 & 0.619 & 0.708 \\  \hline
\end{tabular}  
\end{table}

We next consider the accuracy of the interval estimates produced using multiple imputation for each of the methods. That is, if $\theta$ is the mean of a variable or a regression coefficient, we use Rubin's combining rules \citep{rubin87a} across the multiply imputed datasets to approximate the variance of $\hat\theta^{[r]}(x)$, 
which we denote $T^{[r]}(x)$ at the $r^{\rm th}$ replication. Then, for these parameters, we calculate the coverage of a $(1-\alpha)$\% confidence interval around $\theta$ as $N^{-1}\sum^{N}_{r=1}C^{[r]}_\theta(x)$ where
\begin{equation} \label{cover}
C^{[r]}_\theta(x) =
\left\{
\begin{array}{ll}
1, & \mbox{if $\theta \in \{\hat\theta^{[r]}(x) \pm t_{1-\alpha/2,d^{[r]}}\sqrt{T^{[r]}(x)}\}$}, \\
0, & \mbox{otherwise}.
\end{array}\right.
\end{equation}
and where $t_{\alpha,\nu}$ is the $100\alpha^{\rm th}$ percentile of a $t$ distribution with $\nu$ degrees of freedom (where the degrees of freedom at the $r^{\rm th}$ replication, $d^{[r]}$, are calculated from the within- and between-imputation variances).

\begin{figure}[ht!]
\centering
\includegraphics[width=6.0in, bb = 33 46 762 373]{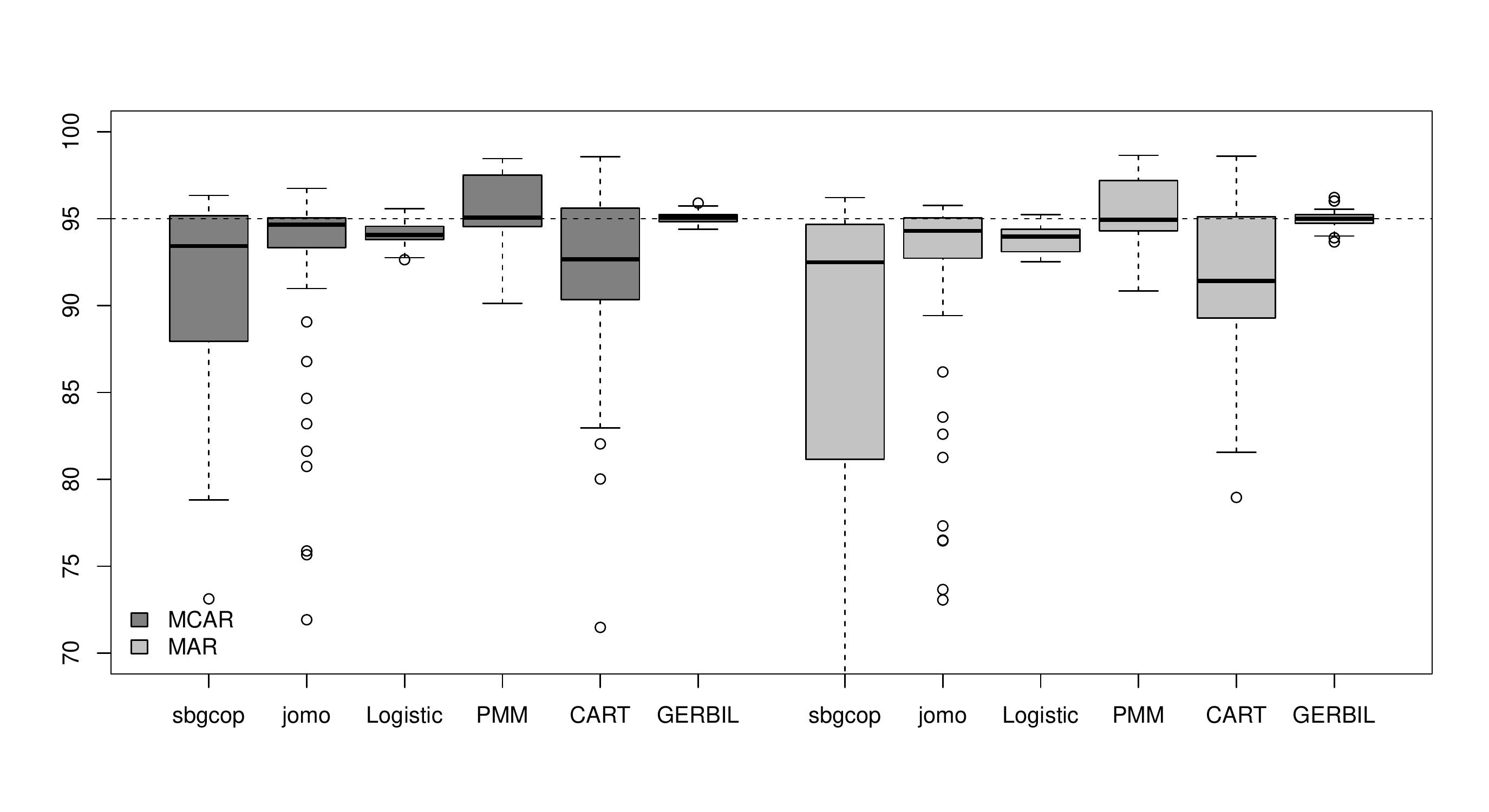}
\caption{Boxplots of the simulated coverage rates for the 95\% confidence intervals of 66 separate parameters under various methods and missingness mechanisms}
\label{figcover}
\end{figure}

Box plots across the 66 parameters for which the coverage rates were calculated are shown in Figure \ref{figcover} for each method and missingness mechanism.  The estimated rates approximate the coverage of a 95\% confidence interval for the parameters.  NMAR results are excluded from the figure since all methods provide poor coverage under NMAR missingness and those results do not further inform the comparative performance of the methods.

Figure \ref{figcover} shows that GERBIL systematically provides estimated coverage that is close to 95\%.  The Logistic and PMM methods perform reasonably well; however, the other methods fail to yield reliable coverage.  Exact rates of coverage are reported in tables provided in Section \ref{simtabs} of the supplemental materials. 

In summary, GERBIL met our aspiration of performing no worse than the procedures the existing procedures within our simulation study. In fact, GERBIL was shown to outperform those methods in several regards.


\section{Discussion} \label{sec5}


The proposed GERBIL method accomplishes the objectives stated at the beginning of Section \ref{sec2}.  Specifically, imputations are sampled from a coherent joint distribution in that the data augmentation framework of \cite{tanner87} is obeyed, there by ensuring MCMC covergence.  Furthermore, it is easily applied in large datasets. Existing general imputation procedures \citep[e.g.,][]{vanbuuren10, su11,  quartagno12} are usually computationally onerous or simply inoperable when applied to high dimensional data. For example, the proposed GERBIL method was upwards of 30 times faster than \texttt{mice} in some applications.
The need for efficient imputation algorithms with high dimensional data is amplified by the fact that big data are becoming increasingly prevalent and that studies have shown the need for exhaustive variable selection when building imputation models \citep{robbins14a}.

GERBIL contains the flexibility to select predictors that are included in each conditional model of the form (\ref{SRmodel}), which is needed for a variety of reasons:
\begin{enumerate}
\item To facilitate the handling of skip logic in that child questions are not allowed to depend upon parent questions; imputation of mixed discrete/continuous variables is often handled in a similar manner \citep[e.g.,][]{robbins13}.
\item To facilitate the nested structure in (\ref{cate}) so that nested indicators of one categorical variable are not allowed to conditionally depend upon one another.
\item To avoid the possiblity of having the number of predictors in the conditional model exceed the number of observed cases for the response variable (or similar restrictions) or to otherwise avoid collinearity issues wherein a predictor matrix would not be of full rank.
\item To invoke expert opinion regarding the interdependence of variables.
\end{enumerate}
However, conditional independence of two variables within models of the form in (\ref{conditionals}) does not imply conditional independence between those variables in models of the form in (\ref{conditionals1}), nor does it imply marginal pairwise independence of the variables.  As such, discretion with the selection of variables for conditional models within GERBIL should perhaps be viewed as a means of obtaining an intuitively and computationally valid joint distribution as opposed to a means by which impermissible variable relationships are prevented.

Similarly, there is potential that (when one is selective with regards to the predictors used within the conditional models) the ordering of the variables may affect the imputations.  However, as the variable ordering used in our data example was natural due to the nearly monotonic nature of the missingness, we did not explore this issue here and leave it for further work.

Although not discussed here directly, the GERBIL procedure can be generalized for use in data of a broader ranger of structures than considered herein.
For instance, the procedure can be reformatted to handle semi-continuous data using the principles outlined by \cite{robbins13}; likewise, it could be reformatted to impute variables that have right-censored observations through the use of a Kaplan-Meier-based transformation. Variables that obey a Poisson (or negative binomial) distribution can be imputed using the empirical distribution transformation (as outlined for continuous variables) or, in a manner that is more theoretically justifiable but also more computationally intensive, imputed using the process outlined here for ordinal variables but while incorporating a Poissonian-type model for the cutpoints.

\bibliographystyle{Chicago}
\bibliography{references}

\newpage

\begin{center}
{\LARGE \bf Supplementary Materials: \\
A Flexible and Efficient Algorithm \\
for Joint Imputation of General Data
} \\
\vspace{.12in}
{\large Michael W.~Robbins}
\end{center}


\renewcommand{\theequation}{A.\arabic{equation}}
\renewcommand{\thetable}{A.\arabic{table}}
\renewcommand{\thefigure}{A.\arabic{figure}}
\renewcommand{\thesection}{A.\arabic{section}}
\setcounter{equation}{0}
\setcounter{figure}{0}
\setcounter{table}{0}
\setcounter{section}{0}







\section{Computational Details of the I Step in GERBIL} \label{dets1}

Here, we illustrate calculation of $\bm{\mu}^{(t)}$ and $\bm{\Sigma}^{(t)}$, the mean vector and covariance matrix of the process $\bm{\psi} = \{\bm{Z}_1,\ldots,\bm{Z}_q\}$ (as described in Section \ref{sec21}) at the $t^{\rm th}$ iteration, from the parameter set $\{ \bm{\beta}^{(t)}_1, \sigma_1^{(t)}, \ldots, \bm{\beta}_q^{(t)}, \sigma_q^{(t)} \}$.  We then illustrate how these quantities are used find the conditional means and variances needed to find the updated version of $\bm{\psi}$ within the I Step of the $t^{\rm th}$ iteration.

For simplicity of illustration, we assume that $\bm{V}_j = \{\bm{1}; \bm{Z}_1; \ldots; \bm{Z}_{j-1}\}$.  As such, (\ref{SRmodel}) becomes
\[ 
\bm{Z}_j =  \beta_{j,0} + {\beta}_{j,1} \bm{Z}_1 + \cdots + \beta_{j,j-1} \bm{Z}_{j-1} + \sigma_j \bm{\epsilon}_j,
\] 
where $\bm{\beta}_j = (\beta_{j,0}, \beta_{j,1},\ldots, \beta_{j,j-1})^T$.

To calculate $\bm{\mu}^{(t)}=\{\mu^{(t)}_{1},\ldots,\mu^{(t)}_{q}\}^T$, set
$
{\mu}^{(t)}_{1}=\beta^{(t)}_{1,0}
$
 and
\[
{\mu}^{(t)}_{j}=\beta^{(t)}_{j,0} + {\beta}^{(t)}_{j,1} \mu^{(t)}_{1} + \cdots + {\beta}^{(t)}_{j,j-1} \mu^{(t)}_{j-1}
\]
sequentially for $j=1,\ldots,q$.

Next, define the following partitions of $\bm{\Sigma}^{(t)}$ for $1\leq \ell \leq q$ and $1 \leq k \leq q$:
{
\begin{eqnarray}
\nonumber {\Sigma}^{(t)}_{\ell,k}~ &-& ~\mbox{The $(\ell,k)^{\rm th}$ element of $\bm{\Sigma}^{(t)}$}, \\
\nonumber \bm{\Sigma}^{(t)}_{1:\ell,k}~ &-& ~\mbox{The first $\ell$ elements of $k^{\rm th}$ column of $\bm{\Sigma}^{(t)}$}, \\
\nonumber \bm{\Sigma}^{(t)}_{\ell,1:k}~ &-& ~\mbox{The first $k$ elements of $\ell^{\rm th}$ row of $\bm{\Sigma}^{(t)}$}, \\
\nonumber \bm{\Sigma}^{(t)}_{1:\ell,1:k}~ &-& ~\mbox{The upper left $\ell \times k$ block of $\bm{\Sigma}^{(t)}$}, \\
\nonumber \bm{\Sigma}^{(t)}_{-\ell,-k}~ &-& ~\mbox{A version of $\bm{\Sigma}^{(t)}$ where the $\ell^{\rm th}$ row and $k^{\rm th}$ column have been removed,} \\
\nonumber \bm{\Sigma}^{(t)}_{-\ell,k}~ &-& ~\mbox{The $k^{\rm th}$ column of $\bm{\Sigma}^{(t)}$ with the $\ell^{\rm th}$ element of that column removed,} \\
\nonumber \bm{\Sigma}^{(t)}_{\ell,-k}~ &-& ~\mbox{The $\ell^{\rm th}$ row of $\bm{\Sigma}^{(t)}$ with the $k^{\rm th}$ element of that row removed.}
\end{eqnarray}
}
We can calculate $\bm{\Sigma}^{(t)}$ sequentially by setting $\Sigma^{(t)}_{1,1}=(\sigma^{(t)}_1)^2$ and
\begin{eqnarray}
\nonumber \bm{\Sigma}^{(t)}_{1:j-1,j} &=& ~(\widetilde{\bm{\beta}}^{(t)}_j)^T\bm{\Sigma}^{(t)}_{1:j-1,1:j-1}, \\
\nonumber \bm{\Sigma}^{(t)}_{j,1:j-1} &=& ~\bm{\Sigma}^{(t)}_{1:j-1,1:j-1}\widetilde{\bm{\beta}}^{(t)}_j, \\
\nonumber {\Sigma}^{(t)}_{j,j} &=& ~(\sigma^{(t)}_j)^2 + (\bm{\beta}^{(t)}_j)^T\bm{\Sigma}^{(t)}_{1:j-1,1:j-1}\widetilde{\bm{\beta}}^{(t)}_j,
\end{eqnarray}
for $j=2,\ldots,p$ and where $\widetilde{\bm{\beta}}_j = (\beta_{j,1},\ldots, \beta_{j,j-1})^T$.

From this, we calculate the quantities
\begin{eqnarray}
\nonumber \mu_{j|\cdot}^{(t+1)} &=& E[\bm{Z}_j|\bm{Z}_1^{(t+1)},\ldots,\bm{Z}_{j-1}^{(t+1)},\bm{Z}_{j+1}^{(t)},\ldots,\bm{Z}_p^{(t)},\bm{\mu}^{(t)},\bm{\Sigma}^{(t)}], \\
\nonumber \sigma_{j|\cdot}^{(t+1)} &=& \mbox{Var}(\bm{Z}_j|\bm{Z}_1^{(t+1)},\ldots,\bm{Z}_{j-1}^{(t+1)},\bm{Z}_{j+1}^{(t)},\ldots,\bm{Z}_p^{(t)},\bm{\mu}^{(t)},\bm{\Sigma}^{(t)}),
\end{eqnarray}
as follows.
%
First, define the length-$(q-1)$ vectors,
\[
\bm{z}_{-j}^{(t+1)} = \{\bm{Z}_{1}^{(t+1)} , \ldots , \bm{Z}_{j-1}^{(t+1)} , \bm{Z}_{j+1}^{(t)} , \ldots , \bm{Z}_{q}^{(t)}\}^{T},
\]
and
\[
\bm{\mu}_{-j}^{(t)} = \{\mu_{1}^{(t)} , \ldots , \mu_{j-1}^{(t)} , \mu_{j+1}^{(t)} , \ldots , \mu_{q}^{(t)}\}^{T},
\]
for each $j=1,\ldots,q$. Then, it follows from multivariate normal theory that
\[ 
\mu_{j|\cdot}^{(t+1)} = \mu_{j}^{(t)} + \bm{\Sigma}_{j,-j}^{(t)} (\bm{\Sigma}_{-j,-j}^{(t)})^{-1} ( \bm{z}_{-j}^{(t+1)} - \bm{\mu}_{-j}^{(t)} ),
\] 
and
\[ 
\sigma_{j|\cdot}^{(t+1)} = \Sigma_{j,j}^{(t)} - { \bm{\Sigma}_{j,-j}^{(t)} } (\bm{\Sigma}_{-j,-j}^{(t)})^{-1} \bm{\Sigma}_{-j,j}^{(t)}.
\] 
Having derived these quantities, $\bm{Z}_j^{(t+1)}$, the updated version of $\bm{Z}_j$, can be calculated in line with the processes outlined in Section \ref{sec32}.
Note that, as mentioned in Section \ref{sec34}, the SWEEP operator 
is used to expedite computation of $\mu_{j|\cdot}^{(t+1)}$ and $\sigma_{j|\cdot}^{(t+1)}$.

\section{Simulation Details} \label{simdets}

A synthetic dataset $\{\bm{X}_1,\ldots,\bm{X}_6\}$ is generated as follows.  The first variable is drawn from a 4-level multinomial distribution with categorical probabilities given by $\{1/4,1/4,1/4,1/4\}$; we let $\bm{X}_1$ denote an $n\times 4$ matrix containing the binary indicators for each category. Next, we generate $\bm{\psi}$ from a 4-dimensional multivariate normal distribution with a mean vector of $\bm{0}$ and a covariance matrix that has ones along the diagonal and 1/2 on each off diagonal element.  Letting $\bm{Z}=\{\bm{Z}_2,\ldots,\bm{Z}_5\}$, we write $\bm{Z}=\bm{X}_1\bm{\gamma}+\bm{\psi}$ where $\gamma=\{1/3,1/5,-1/3,-1/5\}'$.  In addition, we let $\bm{\pi}=\bm{X}_1\bm{\rho}+\bm{\psi}\bm{\xi}$ where $\bm\rho=\{1/3,1/5,-1/3,-1/5\}'$ and $\bm\xi=\{1/2,-1/2,-1/3,1/3\}'$.  We set $\bm{X}_2=\bm{Z}_2$, $\bm{X}_3=\bm{Z}_3$, and $\bm{X}_4=1$ if $\bm{Z}_4 \leq 0$ and $\bm{X}_4 = 0$ otherwise.  In addition,
\[
\bm{X}_5 =
\left\{
\begin{array}{ll}
1, & \mbox{if $\bm{Z}_5 \leq -1.5$}, \\
2, & \mbox{if $\bm{Z}_5 \leq 0$ and $\bm{Z}_5 > -1.5$}, \\
3, & \mbox{if $\bm{Z}_5 \leq 1.5$ and $\bm{Z}_5 > 0$}, \\
4, & \mbox{if $\bm{Z}_5 > 1.5$},
\end{array}\right.
\]
Lastly, $\bm{X}_6$ is sampled from a binomial distribution so that $\mbox{Pr}(\bm{X}_6=1)=1/[1+\exp(-\bm\pi)]$.

Missingness is imposed in $\{\bm{X}_1,\ldots,\bm{X}_6\}$ as follows.  Let $\bm{R}_j$ denote an indicator that is unity if $\bm{X}_j$ is missing and zero otherwise for $j \in \{1,3,\ldots,6\}$.   Further, $\mbox{Pr}(\bm{R}_j=1)=1/[1+\exp(-\beta_{j,0}-\beta_{j,1}\bm{X}_{2}-\beta_{j,2}\bm{X}_j)]$ for $j \in \{1,3,\ldots,6\}$.  For all mechanisms, we set $\beta_{j,0}=\log 2$ to obtain a missingness rate of approximately 1/3 for each variable.  Under MCAR missingness, $\beta_{j,1}=0$ and $\beta_{j,2}=0$ for all $j$.  Likewise, under MAR missingness, $\beta_{1,1}=1/2$, $\beta_{3,1}=1$, $\beta_{4,1}=-1$, $\beta_{5,1}=3/4$, and $\beta_{6,1}=-1/2$ with $\beta_{j,2}=0$ for all $j$, and under NMAR missingness, $\beta_{1,2}=1/2$, $\beta_{3,2}=1$, $\beta_{4,2}=-1$, $\beta_{5,2}=3/4$, and $\beta_{6,2}=-1/2$ with $\beta_{j,1}=0$ for all $j$.

\section{Simulation Tables} \label{simtabs}

Tables \ref{MCAR1}-\ref{NMAR5} list the coverage and rMSE for the parameters studied in the simulations of Section \ref{sec3} under the three different missingness mechanisms.


\begin{singlespace}

\begin{table}[!ht]
\centering
{\small
\renewcommand{\tabcolsep}{.12cm}
\caption{Coverage rates and rMSE for the means of the simulated variables (where $\{\bm{X}_{11},\ldots,\bm{X}_{14}\}$ are binary indicators created from $\bm{X}_1$) across the 5,000 simulated datasets under a MCAR missingness mechanism.  
} \label{MCAR1}
\vspace{.12in}
\begin{tabular}{clccccccccc}
\hline \hline
& & $\bm{X}_{11}$ & $\bm{X}_{12}$ & $\bm{X}_{13}$ & $\bm{X}_{14}$ & $\bm{X}_{2}$ & $\bm{X}_{3}$ & $\bm{X}_{4}$ & $\bm{X}_{5}$ & $\bm{X}_{6}$ \\  \hline
\multirow{6}{*}{Coverage} & sbgcop & 0.5870 & 0.5916 & 0.5436 & 0.6476 & --- & 0.8328 & 0.7882 & 0.8794 & 0.8920 \\
 & jomo & 0.9530 & 0.9520 & 0.9528 & 0.9512 & --- & 0.9528 & 0.9504 & 0.9538 & 0.9466 \\
 & Logistic & 0.9372 & 0.9354 & 0.9388 & 0.9374 & --- & 0.9558 & 0.9530 & 0.9378 & 0.9474 \\
 & PMM & 0.9542 & 0.9392 & 0.9418 & 0.9506 & --- & 0.9552 & 0.9518 & 0.9502 & 0.9484 \\
 & CART & 0.9392 & 0.9360 & 0.9376 & 0.9386 & --- & 0.9418 & 0.9388 & 0.9354 & 0.9314 \\
 & GERBIL & 0.9540 & 0.9466 & 0.9522 & 0.9524 & --- & 0.9554 & 0.9510 & 0.9490 & 0.9460 \\  \hline
\multirow{6}{*}{rMSE} & sbgcop & 0.0221 & 0.0231 & 0.0236 & 0.0226 & --- & 0.0374 & 0.0192 & 0.0238 & 0.0157 \\
 & jomo & 0.0117 & 0.0118 & 0.0118 & 0.0115 & --- & 0.0272 & 0.0130 & 0.0196 & 0.0137 \\
 & Logistic & 0.0117 & 0.0119 & 0.0118 & 0.0115 & --- & 0.0271 & 0.0130 & 0.0196 & 0.0137 \\
 & PMM & 0.0118 & 0.0119 & 0.0119 & 0.0116 & --- & 0.0271 & 0.0131 & 0.0197 & 0.0137 \\
 & CART & 0.0118 & 0.0119 & 0.0119 & 0.0116 & --- & 0.0274 & 0.0132 & 0.0198 & 0.0139 \\
 & GERBIL & 0.0117 & 0.0119 & 0.0118 & 0.0115 & --- & 0.0271 & 0.0131 & 0.0197 & 0.0138 \\  \hline
\end{tabular}
}
\end{table}

\begin{table}[!ht]
\centering
{
\footnotesize
\renewcommand{\tabcolsep}{.12cm}
\caption{\footnotesize Coverage rates using six methods of imputation for the parameters of the fully-specified regression models of the simulated variables (where $\{\bm{X}_{12},\ldots,\bm{X}_{14}\}$ are binary indicators created from $\bm{X}_1$) across the 5,000 simulated datasets under a MCAR missingness mechanism. Rows indicate the outcome variable and columns indicate predictors.  
} \label{MCAR2}
\vspace{.12in}
\begin{tabular}{cccccccccccc}
\hline \hline
 & & Intercept & $\bm{X}_{12}$ & $\bm{X}_{13}$ & $\bm{X}_{14}$ & $\bm{X}_{2}$ & $\bm{X}_{3}$ & $\bm{X}_{4}$ & $\bm{X}_{5}$ & $\bm{X}_{6}$ \\  \hline
\multirow{8}{*}{sbgcop} & $\bm{X}_{12}$ & 0.9586 & --- & --- & --- & 0.9542 & 0.9622 & 0.9276 & 0.9500 & 0.9386 \\
 & $\bm{X}_{13}$ & 0.9610 & --- & --- & --- & 0.9536 & 0.9610 & 0.9152 & 0.9530 & 0.9330 \\
 & $\bm{X}_{14}$ & 0.9474 & --- & --- & --- & 0.9586 & 0.9616 & 0.9254 & 0.9498 & 0.9346 \\
 & $\bm{X}_{2}$ & 0.9156 & 0.9538 & 0.9532 & 0.9586 & --- & 0.8788 & 0.9482 & 0.9066 & 0.9432 \\
 & $\bm{X}_{3}$ & 0.7904 & 0.9634 & 0.9608 & 0.9620 & 0.8722 & --- & 0.9518 & 0.8828 & 0.9300 \\
 & $\bm{X}_{4}$ & 0.7928 & 0.9272 & 0.9196 & 0.9288 & 0.9432 & 0.9434 & --- & 0.7312 & 0.9360 \\
 & $\bm{X}_{5}$ & 0.8256 & 0.9512 & 0.9520 & 0.9486 & 0.9160 & 0.8790 & 0.6764 & --- & 0.8032 \\
 & $\bm{X}_{6}$ & 0.8814 & 0.9392 & 0.9330 & 0.9340 & 0.9408 & 0.9444 & 0.9356 & 0.8272 & --- \\  \hline
\multirow{8}{*}{jomo} & $\bm{X}_{12}$ & 0.9248 & --- & --- & --- & 0.9524 & 0.9514 & 0.9462 & 0.9146 & 0.9492 \\
 & $\bm{X}_{13}$ & 0.9406 & --- & --- & --- & 0.9514 & 0.9506 & 0.9468 & 0.9334 & 0.9466 \\
 & $\bm{X}_{14}$ & 0.7566 & --- & --- & --- & 0.9464 & 0.9476 & 0.9516 & 0.7192 & 0.9434 \\
 & $\bm{X}_{2}$ & 0.9616 & 0.9502 & 0.9500 & 0.9456 & --- & 0.9340 & 0.9438 & 0.9674 & 0.9506 \\
 & $\bm{X}_{3}$ & 0.8320 & 0.9478 & 0.9502 & 0.9450 & 0.9178 & --- & 0.9306 & 0.7588 & 0.9496 \\
 & $\bm{X}_{4}$ & 0.8466 & 0.9476 & 0.9480 & 0.9500 & 0.9482 & 0.9438 & --- & 0.8074 & 0.9504 \\
 & $\bm{X}_{5}$ & 0.8678 & 0.9384 & 0.9414 & 0.8162 & 0.9356 & 0.8906 & 0.9098 & --- & 0.9384 \\
 & $\bm{X}_{6}$ & 0.9330 & 0.9486 & 0.9476 & 0.9436 & 0.9532 & 0.9456 & 0.9480 & 0.9140 & --- \\  \hline
\multirow{8}{*}{Logistic} & $\bm{X}_{12}$ & 0.9372 & --- & --- & --- & 0.9380 & 0.9426 & 0.9414 & 0.9326 & 0.9406 \\
 & $\bm{X}_{13}$ & 0.9330 & --- & --- & --- & 0.9380 & 0.9456 & 0.9394 & 0.9326 & 0.9386 \\
 & $\bm{X}_{14}$ & 0.9322 & --- & --- & --- & 0.9386 & 0.9462 & 0.9386 & 0.9276 & 0.9406 \\
 & $\bm{X}_{2}$ & 0.9416 & 0.9386 & 0.9360 & 0.9420 & --- & 0.9420 & 0.9484 & 0.9428 & 0.9488 \\
 & $\bm{X}_{3}$ & 0.9476 & 0.9426 & 0.9426 & 0.9440 & 0.9440 & --- & 0.9522 & 0.9428 & 0.9462 \\
 & $\bm{X}_{4}$ & 0.9402 & 0.9410 & 0.9410 & 0.9402 & 0.9468 & 0.9520 & --- & 0.9388 & 0.9484 \\
 & $\bm{X}_{5}$ & 0.9404 & 0.9292 & 0.9292 & 0.9264 & 0.9400 & 0.9408 & 0.9370 & --- & 0.9406 \\
 & $\bm{X}_{6}$ & 0.9478 & 0.9404 & 0.9374 & 0.9400 & 0.9504 & 0.9462 & 0.9492 & 0.9436 & --- \\  \hline
\multirow{8}{*}{PMM} & $\bm{X}_{12}$ & 0.9728 & --- & --- & --- & 0.9802 & 0.9758 & 0.9802 & 0.9802 & 0.9170 \\
 & $\bm{X}_{13}$ & 0.9788 & --- & --- & --- & 0.9814 & 0.9780 & 0.9830 & 0.9800 & 0.9460 \\
 & $\bm{X}_{14}$ & 0.9388 & --- & --- & --- & 0.9506 & 0.9434 & 0.9344 & 0.9376 & 0.9012 \\
 & $\bm{X}_{2}$ & 0.9542 & 0.9800 & 0.9794 & 0.9472 & --- & 0.9416 & 0.9480 & 0.9530 & 0.9458 \\
 & $\bm{X}_{3}$ & 0.9542 & 0.9750 & 0.9768 & 0.9412 & 0.9506 & --- & 0.9562 & 0.9478 & 0.9458 \\
 & $\bm{X}_{4}$ & 0.9446 & 0.9810 & 0.9846 & 0.9368 & 0.9494 & 0.9540 & --- & 0.9466 & 0.9506 \\
 & $\bm{X}_{5}$ & 0.9554 & 0.9798 & 0.9782 & 0.9356 & 0.9496 & 0.9492 & 0.9432 & --- & 0.9474 \\
 & $\bm{X}_{6}$ & 0.9550 & 0.9144 & 0.9454 & 0.9018 & 0.9518 & 0.9464 & 0.9508 & 0.9486 & --- \\  \hline
\multirow{8}{*}{CART} & $\bm{X}_{12}$ & 0.9218 & --- & --- & --- & 0.9178 & 0.9654 & 0.9720 & 0.9178 & 0.9696 \\
 & $\bm{X}_{13}$ & 0.9748 & --- & --- & --- & 0.9500 & 0.9684 & 0.9850 & 0.9642 & 0.8628 \\
 & $\bm{X}_{14}$ & 0.8632 & --- & --- & --- & 0.9214 & 0.9598 & 0.9582 & 0.8364 & 0.9184 \\
 & $\bm{X}_{2}$ & 0.9032 & 0.9182 & 0.9494 & 0.9196 & --- & 0.9224 & 0.9326 & 0.9150 & 0.9364 \\
 & $\bm{X}_{3}$ & 0.9128 & 0.9642 & 0.9668 & 0.9580 & 0.9240 & --- & 0.8794 & 0.9034 & 0.9100 \\
 & $\bm{X}_{4}$ & 0.8308 & 0.9704 & 0.9856 & 0.9560 & 0.9314 & 0.8874 & --- & 0.8002 & 0.9180 \\
 & $\bm{X}_{5}$ & 0.7148 & 0.9294 & 0.9688 & 0.8614 & 0.8800 & 0.9132 & 0.8204 & --- & 0.8504 \\
 & $\bm{X}_{6}$ & 0.9380 & 0.9680 & 0.8592 & 0.9176 & 0.9386 & 0.9030 & 0.9146 & 0.8296 & --- \\  \hline
\multirow{8}{*}{GERBIL} & $\bm{X}_{12}$ & 0.9590 & --- & --- & --- & 0.9502 & 0.9554 & 0.9522 & 0.9568 & 0.9524 \\
 & $\bm{X}_{13}$ & 0.9524 & --- & --- & --- & 0.9486 & 0.9534 & 0.9496 & 0.9510 & 0.9486 \\
 & $\bm{X}_{14}$ & 0.9512 & --- & --- & --- & 0.9574 & 0.9568 & 0.9488 & 0.9474 & 0.9516 \\
 & $\bm{X}_{2}$ & 0.9504 & 0.9498 & 0.9474 & 0.9552 & --- & 0.9462 & 0.9478 & 0.9534 & 0.9500 \\
 & $\bm{X}_{3}$ & 0.9514 & 0.9530 & 0.9526 & 0.9542 & 0.9516 & --- & 0.9520 & 0.9476 & 0.9482 \\
 & $\bm{X}_{4}$ & 0.9454 & 0.9518 & 0.9476 & 0.9474 & 0.9484 & 0.9514 & --- & 0.9504 & 0.9470 \\
 & $\bm{X}_{5}$ & 0.9544 & 0.9540 & 0.9472 & 0.9440 & 0.9472 & 0.9518 & 0.9482 & --- & 0.9462 \\
 & $\bm{X}_{6}$ & 0.9550 & 0.9522 & 0.9494 & 0.9514 & 0.9490 & 0.9480 & 0.9482 & 0.9482 & --- \\  \hline
\end{tabular}
}
\end{table}

\begin{table}[!ht]
\centering
{
\footnotesize
\renewcommand{\tabcolsep}{.12cm}
\caption{\footnotesize rMSE using six methods of imputation for the parameters of the fully-specified regression models of the simulated variables (where $\{\bm{X}_{12},\ldots,\bm{X}_{14}\}$ are binary indicators created from $\bm{X}_1$) across the 5,000 simulated datasets under a MCAR missingness mechanism. Rows indicate the outcome variable and columns indicate predictors.  
} \label{MCAR3}
\vspace{.12in}
\begin{tabular}{cccccccccccc}
\hline \hline
 & & Intercept & $\bm{X}_{12}$ & $\bm{X}_{13}$ & $\bm{X}_{14}$ & $\bm{X}_{2}$ & $\bm{X}_{3}$ & $\bm{X}_{4}$ & $\bm{X}_{5}$ & $\bm{X}_{6}$ \\  \hline
\multirow{8}{*}{sbgcop} & $\bm{X}_{12}$ & 0.4032 & --- & --- & --- & 0.1081 & 0.1215 & 0.2662 & 0.1713 & 0.2302 \\
 & $\bm{X}_{13}$ & 0.3794 & --- & --- & --- & 0.1030 & 0.1177 & 0.2714 & 0.1659 & 0.2242 \\
 & $\bm{X}_{14}$ & 0.4434 & --- & --- & --- & 0.1088 & 0.1239 & 0.2782 & 0.1811 & 0.2396 \\
 & $\bm{X}_{2}$ & 0.1109 & 0.0682 & 0.0654 & 0.0681 & --- & 0.0338 & 0.0568 & 0.0457 & 0.0496 \\
 & $\bm{X}_{3}$ & 0.1713 & 0.0761 & 0.0748 & 0.0776 & 0.0355 & --- & 0.0654 & 0.0564 & 0.0602 \\
 & $\bm{X}_{4}$ & 0.5737 & 0.2689 & 0.2736 & 0.2806 & 0.0947 & 0.1090 & --- & 0.2369 & 0.1960 \\
 & $\bm{X}_{5}$ & 0.0751 & 0.0611 & 0.0593 & 0.0644 & 0.0256 & 0.0316 & 0.0860 & --- & 0.0648 \\
 & $\bm{X}_{6}$ & 0.3967 & 0.2293 & 0.2231 & 0.2392 & 0.0818 & 0.0968 & 0.1920 & 0.1797 & --- \\  \hline
\multirow{8}{*}{jomo} & $\bm{X}_{12}$ & 0.4512 & --- & --- & --- & 0.1121 & 0.1310 & 0.2392 & 0.1834 & 0.2145 \\
 & $\bm{X}_{13}$ & 0.4043 & --- & --- & --- & 0.1067 & 0.1243 & 0.2392 & 0.1626 & 0.2075 \\
 & $\bm{X}_{14}$ & 0.6806 & --- & --- & --- & 0.1159 & 0.1368 & 0.2517 & 0.2759 & 0.2229 \\
 & $\bm{X}_{2}$ & 0.0901 & 0.0698 & 0.0670 & 0.0719 & --- & 0.0286 & 0.0557 & 0.0342 & 0.0482 \\
 & $\bm{X}_{3}$ & 0.1634 & 0.0825 & 0.0789 & 0.0867 & 0.0321 & --- & 0.0686 & 0.0700 & 0.0569 \\
 & $\bm{X}_{4}$ & 0.5007 & 0.2411 & 0.2423 & 0.2541 & 0.0935 & 0.1107 & --- & 0.1956 & 0.1763 \\
 & $\bm{X}_{5}$ & 0.0717 & 0.0637 & 0.0606 & 0.0914 & 0.0239 & 0.0317 & 0.0578 & --- & 0.0449 \\
 & $\bm{X}_{6}$ & 0.3258 & 0.2139 & 0.2068 & 0.2226 & 0.0802 & 0.0937 & 0.1741 & 0.1276 & --- \\  \hline
\multirow{8}{*}{Logistic} & $\bm{X}_{12}$ & 0.4306 & --- & --- & --- & 0.1124 & 0.1327 & 0.2408 & 0.1736 & 0.2178 \\
 & $\bm{X}_{13}$ & 0.4124 & --- & --- & --- & 0.1073 & 0.1263 & 0.2400 & 0.1687 & 0.2109 \\
 & $\bm{X}_{14}$ & 0.4585 & --- & --- & --- & 0.1146 & 0.1373 & 0.2559 & 0.1832 & 0.2268 \\
 & $\bm{X}_{2}$ & 0.0961 & 0.0699 & 0.0673 & 0.0709 & --- & 0.0274 & 0.0549 & 0.0380 & 0.0479 \\
 & $\bm{X}_{3}$ & 0.1155 & 0.0820 & 0.0789 & 0.0849 & 0.0284 & --- & 0.0644 & 0.0439 & 0.0566 \\
 & $\bm{X}_{4}$ & 0.3849 & 0.2420 & 0.2417 & 0.2573 & 0.0921 & 0.1074 & --- & 0.1417 & 0.1778 \\
 & $\bm{X}_{5}$ & 0.0533 & 0.0614 & 0.0596 & 0.0646 & 0.0220 & 0.0245 & 0.0496 & --- & 0.0434 \\
 & $\bm{X}_{6}$ & 0.3175 & 0.2172 & 0.2105 & 0.2265 & 0.0799 & 0.0950 & 0.1766 & 0.1243 & --- \\  \hline
\multirow{8}{*}{PMM} & $\bm{X}_{12}$ & 0.3890 & --- & --- & --- & 0.0933 & 0.1149 & 0.2031 & 0.1484 & 0.2503 \\
 & $\bm{X}_{13}$ & 0.3383 & --- & --- & --- & 0.0858 & 0.1014 & 0.1881 & 0.1369 & 0.2053 \\
 & $\bm{X}_{14}$ & 0.4960 & --- & --- & --- & 0.1202 & 0.1476 & 0.2745 & 0.1962 & 0.2747 \\
 & $\bm{X}_{2}$ & 0.0955 & 0.0580 & 0.0538 & 0.0744 & --- & 0.0276 & 0.0548 & 0.0384 & 0.0483 \\
 & $\bm{X}_{3}$ & 0.1143 & 0.0714 & 0.0636 & 0.0916 & 0.0286 & --- & 0.0636 & 0.0443 & 0.0573 \\
 & $\bm{X}_{4}$ & 0.3899 & 0.2035 & 0.1886 & 0.2757 & 0.0912 & 0.1055 & --- & 0.1451 & 0.1759 \\
 & $\bm{X}_{5}$ & 0.0526 & 0.0522 & 0.0480 & 0.0686 & 0.0221 & 0.0244 & 0.0503 & --- & 0.0435 \\
 & $\bm{X}_{6}$ & 0.3149 & 0.2504 & 0.2052 & 0.2747 & 0.0805 & 0.0954 & 0.1749 & 0.1262 & --- \\  \hline
\multirow{8}{*}{CART} & $\bm{X}_{12}$ & 0.4458 & --- & --- & --- & 0.1201 & 0.1118 & 0.1966 & 0.1744 & 0.1754 \\
 & $\bm{X}_{13}$ & 0.3255 & --- & --- & --- & 0.1013 & 0.1025 & 0.1676 & 0.1366 & 0.2455 \\
 & $\bm{X}_{14}$ & 0.5409 & --- & --- & --- & 0.1237 & 0.1209 & 0.2202 & 0.2206 & 0.2276 \\
 & $\bm{X}_{2}$ & 0.1083 & 0.0731 & 0.0628 & 0.0745 & --- & 0.0287 & 0.0556 & 0.0398 & 0.0476 \\
 & $\bm{X}_{3}$ & 0.1254 & 0.0696 & 0.0646 & 0.0753 & 0.0301 & --- & 0.0741 & 0.0478 & 0.0603 \\
 & $\bm{X}_{4}$ & 0.5048 & 0.1985 & 0.1686 & 0.2230 & 0.0944 & 0.1231 & --- & 0.1964 & 0.1803 \\
 & $\bm{X}_{5}$ & 0.0821 & 0.0610 & 0.0486 & 0.0766 & 0.0268 & 0.0268 & 0.0676 & --- & 0.0525 \\
 & $\bm{X}_{6}$ & 0.3105 & 0.1749 & 0.2453 & 0.2272 & 0.0786 & 0.0990 & 0.1781 & 0.1504 & --- \\  \hline
\multirow{8}{*}{GERBIL} & $\bm{X}_{12}$ & 0.4281 & --- & --- & --- & 0.1118 & 0.1317 & 0.2410 & 0.1730 & 0.2170 \\
 & $\bm{X}_{13}$ & 0.4104 & --- & --- & --- & 0.1071 & 0.1261 & 0.2402 & 0.1686 & 0.2100 \\
 & $\bm{X}_{14}$ & 0.4524 & --- & --- & --- & 0.1133 & 0.1352 & 0.2541 & 0.1812 & 0.2236 \\
 & $\bm{X}_{2}$ & 0.0966 & 0.0695 & 0.0670 & 0.0702 & --- & 0.0274 & 0.0550 & 0.0382 & 0.0481 \\
 & $\bm{X}_{3}$ & 0.1158 & 0.0810 & 0.0783 & 0.0834 & 0.0283 & --- & 0.0644 & 0.0442 & 0.0572 \\
 & $\bm{X}_{4}$ & 0.3916 & 0.2440 & 0.2446 & 0.2572 & 0.0927 & 0.1094 & --- & 0.1441 & 0.1818 \\
 & $\bm{X}_{5}$ & 0.0529 & 0.0607 & 0.0590 & 0.0636 & 0.0220 & 0.0246 & 0.0495 & --- & 0.0439 \\
 & $\bm{X}_{6}$ & 0.3225 & 0.2161 & 0.2091 & 0.2232 & 0.0804 & 0.0969 & 0.1784 & 0.1269 & --- \\  \hline
\end{tabular}
}
\end{table}

\begin{table}[!ht]
\centering
{
\scriptsize
\renewcommand{\tabcolsep}{.12cm}
\caption{\footnotesize rMSE using six methods of imputation for elements of the variance-covariance matrix calculated with the simulated variables (where $\{\bm{X}_{11},\ldots,\bm{X}_{14}\}$ are binary indicators created from $\bm{X}_1$) across the 5,000 simulated datasets under a MCAR missingness mechanism. 
} \label{MCAR4}
\vspace{.12in}
\begin{tabular}{ccccccccccccc}
\hline \hline
 & & $\bm{X}_{11}$ & $\bm{X}_{12}$ & $\bm{X}_{13}$ & $\bm{X}_{14}$ & $\bm{X}_{2}$ & $\bm{X}_{3}$ & $\bm{X}_{4}$ & $\bm{X}_{5}$ & $\bm{X}_{6}$ \\  \hline
\multirow{9}{*}{sbgcop} & $\bm{X}_{11}$ & 0.0110 & --- & --- & --- & --- & --- & --- & --- & --- \\
 & $\bm{X}_{12}$ & 0.0065 & 0.0114 & --- & --- & --- & --- & --- & --- & --- \\
 & $\bm{X}_{13}$ & 0.0067 & 0.0064 & 0.0119 & --- & --- & --- & --- & --- & --- \\
 & $\bm{X}_{14}$ & 0.0064 & 0.0068 & 0.0065 & 0.0109 & --- & --- & --- & --- & --- \\
 & $\bm{X}_{2}$ & 0.0130 & 0.0129 & 0.0125 & 0.0133 & --- & --- & --- & --- & --- \\
 & $\bm{X}_{3}$ & 0.0146 & 0.0144 & 0.0143 & 0.0148 & 0.0327 & 0.0429 & --- & --- & --- \\
 & $\bm{X}_{4}$ & 0.0070 & 0.0074 & 0.0072 & 0.0074 & 0.0123 & 0.0144 & 0.0005 & --- & --- \\
 & $\bm{X}_{5}$ & 0.0106 & 0.0110 & 0.0101 & 0.0122 & 0.0265 & 0.0347 & 0.0121 & 0.0200 & --- \\
 & $\bm{X}_{6}$ & 0.0075 & 0.0078 & 0.0074 & 0.0075 & 0.0143 & 0.0165 & 0.0086 & 0.0166 & 0.0004 \\  \hline
\multirow{9}{*}{jomo} & $\bm{X}_{11}$ & 0.0058 & --- & --- & --- & --- & --- & --- & --- & --- \\
 & $\bm{X}_{12}$ & 0.0034 & 0.0059 & --- & --- & --- & --- & --- & --- & --- \\
 & $\bm{X}_{13}$ & 0.0033 & 0.0034 & 0.0059 & --- & --- & --- & --- & --- & --- \\
 & $\bm{X}_{14}$ & 0.0034 & 0.0033 & 0.0034 & 0.0058 & --- & --- & --- & --- & --- \\
 & $\bm{X}_{2}$ & 0.0126 & 0.0123 & 0.0124 & 0.0135 & --- & --- & --- & --- & --- \\
 & $\bm{X}_{3}$ & 0.0146 & 0.0140 & 0.0142 & 0.0161 & 0.0332 & 0.0433 & --- & --- & --- \\
 & $\bm{X}_{4}$ & 0.0070 & 0.0070 & 0.0070 & 0.0076 & 0.0126 & 0.0150 & 0.0003 & --- & --- \\
 & $\bm{X}_{5}$ & 0.0147 & 0.0094 & 0.0118 & 0.0191 & 0.0306 & 0.0459 & 0.0193 & 0.0198 & --- \\
 & $\bm{X}_{6}$ & 0.0071 & 0.0071 & 0.0071 & 0.0070 & 0.0144 & 0.0166 & 0.0081 & 0.0113 & 0.0003 \\  \hline
\multirow{9}{*}{Logistic} & $\bm{X}_{11}$ & 0.0059 & --- & --- & --- & --- & --- & --- & --- & --- \\
 & $\bm{X}_{12}$ & 0.0034 & 0.0059 & --- & --- & --- & --- & --- & --- & --- \\
 & $\bm{X}_{13}$ & 0.0033 & 0.0034 & 0.0059 & --- & --- & --- & --- & --- & --- \\
 & $\bm{X}_{14}$ & 0.0034 & 0.0033 & 0.0034 & 0.0058 & --- & --- & --- & --- & --- \\
 & $\bm{X}_{2}$ & 0.0123 & 0.0122 & 0.0123 & 0.0123 & --- & --- & --- & --- & --- \\
 & $\bm{X}_{3}$ & 0.0141 & 0.0140 & 0.0141 & 0.0143 & 0.0327 & 0.0430 & --- & --- & --- \\
 & $\bm{X}_{4}$ & 0.0067 & 0.0069 & 0.0069 & 0.0068 & 0.0122 & 0.0141 & 0.0003 & --- & --- \\
 & $\bm{X}_{5}$ & 0.0101 & 0.0100 & 0.0101 & 0.0101 & 0.0222 & 0.0248 & 0.0105 & 0.0199 & --- \\
 & $\bm{X}_{6}$ & 0.0071 & 0.0072 & 0.0072 & 0.0071 & 0.0144 & 0.0166 & 0.0081 & 0.0118 & 0.0003 \\  \hline
\multirow{9}{*}{PMM} & $\bm{X}_{11}$ & 0.0059 & --- & --- & --- & --- & --- & --- & --- & --- \\
 & $\bm{X}_{12}$ & 0.0034 & 0.0059 & --- & --- & --- & --- & --- & --- & --- \\
 & $\bm{X}_{13}$ & 0.0034 & 0.0034 & 0.0059 & --- & --- & --- & --- & --- & --- \\
 & $\bm{X}_{14}$ & 0.0034 & 0.0034 & 0.0034 & 0.0058 & --- & --- & --- & --- & --- \\
 & $\bm{X}_{2}$ & 0.0123 & 0.0122 & 0.0129 & 0.0125 & --- & --- & --- & --- & --- \\
 & $\bm{X}_{3}$ & 0.0138 & 0.0130 & 0.0139 & 0.0143 & 0.0327 & 0.0429 & --- & --- & --- \\
 & $\bm{X}_{4}$ & 0.0066 & 0.0058 & 0.0060 & 0.0067 & 0.0122 & 0.0141 & 0.0003 & --- & --- \\
 & $\bm{X}_{5}$ & 0.0100 & 0.0091 & 0.0096 & 0.0102 & 0.0222 & 0.0249 & 0.0106 & 0.0199 & --- \\
 & $\bm{X}_{6}$ & 0.0069 & 0.0084 & 0.0103 & 0.0076 & 0.0144 & 0.0166 & 0.0080 & 0.0118 & 0.0003 \\  \hline
\multirow{9}{*}{CART} & $\bm{X}_{11}$ & 0.0059 & --- & --- & --- & --- & --- & --- & --- & --- \\
 & $\bm{X}_{12}$ & 0.0035 & 0.0060 & --- & --- & --- & --- & --- & --- & --- \\
 & $\bm{X}_{13}$ & 0.0034 & 0.0034 & 0.0059 & --- & --- & --- & --- & --- & --- \\
 & $\bm{X}_{14}$ & 0.0034 & 0.0034 & 0.0035 & 0.0058 & --- & --- & --- & --- & --- \\
 & $\bm{X}_{2}$ & 0.0124 & 0.0123 & 0.0123 & 0.0123 & --- & --- & --- & --- & --- \\
 & $\bm{X}_{3}$ & 0.0145 & 0.0141 & 0.0135 & 0.0148 & 0.0330 & 0.0436 & --- & --- & --- \\
 & $\bm{X}_{4}$ & 0.0069 & 0.0064 & 0.0060 & 0.0070 & 0.0124 & 0.0152 & 0.0003 & --- & --- \\
 & $\bm{X}_{5}$ & 0.0123 & 0.0094 & 0.0086 & 0.0115 & 0.0226 & 0.0268 & 0.0146 & 0.0200 & --- \\
 & $\bm{X}_{6}$ & 0.0079 & 0.0066 & 0.0081 & 0.0069 & 0.0146 & 0.0176 & 0.0081 & 0.0122 & 0.0003 \\  \hline
\multirow{9}{*}{GERBIL} & $\bm{X}_{11}$ & 0.0059 & --- & --- & --- & --- & --- & --- & --- & --- \\
 & $\bm{X}_{12}$ & 0.0034 & 0.0060 & --- & --- & --- & --- & --- & --- & --- \\
 & $\bm{X}_{13}$ & 0.0034 & 0.0034 & 0.0059 & --- & --- & --- & --- & --- & --- \\
 & $\bm{X}_{14}$ & 0.0034 & 0.0033 & 0.0034 & 0.0058 & --- & --- & --- & --- & --- \\
 & $\bm{X}_{2}$ & 0.0119 & 0.0120 & 0.0121 & 0.0119 & --- & --- & --- & --- & --- \\
 & $\bm{X}_{3}$ & 0.0136 & 0.0137 & 0.0138 & 0.0138 & 0.0326 & 0.0430 & --- & --- & --- \\
 & $\bm{X}_{4}$ & 0.0066 & 0.0069 & 0.0069 & 0.0067 & 0.0122 & 0.0141 & 0.0003 & --- & --- \\
 & $\bm{X}_{5}$ & 0.0098 & 0.0099 & 0.0100 & 0.0099 & 0.0222 & 0.0249 & 0.0105 & 0.0199 & --- \\
 & $\bm{X}_{6}$ & 0.0071 & 0.0072 & 0.0072 & 0.0070 & 0.0144 & 0.0166 & 0.0081 & 0.0119 & 0.0003 \\  \hline
\end{tabular}
}
\end{table}

\begin{table}[!ht]
\centering
{
\footnotesize
\renewcommand{\tabcolsep}{.12cm}
\caption{\footnotesize rMSE using six methods of imputation for the squared standard errors of the parameters of the fully-specified regression models with the simulated variables (where $\{\bm{X}_{12},\ldots,\bm{X}_{14}\}$ are binary indicators created from $\bm{X}_1$) across the 5,000 simulated datasets under a MCAR missingness mechanism. Rows indicate the outcome variable and columns indicate predictors.  
} \label{MCAR5}
\vspace{.12in}
\begin{tabular}{cccccccccccc}
\hline \hline
 & & Intercept & $\bm{X}_{12}$ & $\bm{X}_{13}$ & $\bm{X}_{14}$ & $\bm{X}_{2}$ & $\bm{X}_{3}$ & $\bm{X}_{4}$ & $\bm{X}_{5}$ & $\bm{X}_{6}$ \\  \hline
\multirow{8}{*}{sbgcop} & $\bm{X}_{12}$ & 0.2087 & --- & --- & --- & 0.0775 & 0.0776 & 0.1301 & 0.0997 & 0.1196 \\
 & $\bm{X}_{13}$ & 0.1998 & --- & --- & --- & 0.0743 & 0.0744 & 0.1287 & 0.0963 & 0.1153 \\
 & $\bm{X}_{14}$ & 0.2160 & --- & --- & --- & 0.0806 & 0.0805 & 0.1345 & 0.1035 & 0.1235 \\
 & $\bm{X}_{2}$ & 0.0705 & 0.0501 & 0.0484 & 0.0519 & --- & 0.0207 & 0.0400 & 0.0283 & 0.0352 \\
 & $\bm{X}_{3}$ & 0.0721 & 0.0500 & 0.0483 & 0.0519 & 0.0206 & --- & 0.0401 & 0.0279 & 0.0351 \\
 & $\bm{X}_{4}$ & 0.1857 & 0.1307 & 0.1294 & 0.1352 & 0.0646 & 0.0644 & --- & 0.0821 & 0.0994 \\
 & $\bm{X}_{5}$ & 0.0339 & 0.0381 & 0.0367 & 0.0395 & 0.0160 & 0.0159 & 0.0304 & --- & 0.0270 \\
 & $\bm{X}_{6}$ & 0.1658 & 0.1194 & 0.1151 & 0.1234 & 0.0560 & 0.0563 & 0.0987 & 0.0724 & --- \\  \hline
\multirow{8}{*}{jomo} & $\bm{X}_{12}$ & 0.2088 & --- & --- & --- & 0.0775 & 0.0777 & 0.1310 & 0.1007 & 0.1201 \\
 & $\bm{X}_{13}$ & 0.2000 & --- & --- & --- & 0.0744 & 0.0745 & 0.1297 & 0.0972 & 0.1159 \\
 & $\bm{X}_{14}$ & 0.2177 & --- & --- & --- & 0.0805 & 0.0805 & 0.1355 & 0.1045 & 0.1240 \\
 & $\bm{X}_{2}$ & 0.0706 & 0.0502 & 0.0485 & 0.0519 & --- & 0.0207 & 0.0401 & 0.0283 & 0.0353 \\
 & $\bm{X}_{3}$ & 0.0720 & 0.0501 & 0.0484 & 0.0519 & 0.0205 & --- & 0.0401 & 0.0280 & 0.0351 \\
 & $\bm{X}_{4}$ & 0.1905 & 0.1316 & 0.1306 & 0.1362 & 0.0647 & 0.0645 & --- & 0.0832 & 0.1001 \\
 & $\bm{X}_{5}$ & 0.0339 & 0.0380 & 0.0367 & 0.0394 & 0.0160 & 0.0159 & 0.0304 & --- & 0.0269 \\
 & $\bm{X}_{6}$ & 0.1664 & 0.1198 & 0.1157 & 0.1239 & 0.0560 & 0.0564 & 0.0992 & 0.0729 & --- \\  \hline
\multirow{8}{*}{Logistic} & $\bm{X}_{12}$ & 0.2049 & --- & --- & --- & 0.0774 & 0.0775 & 0.1305 & 0.0996 & 0.1199 \\
 & $\bm{X}_{13}$ & 0.1972 & --- & --- & --- & 0.0744 & 0.0744 & 0.1293 & 0.0963 & 0.1158 \\
 & $\bm{X}_{14}$ & 0.2107 & --- & --- & --- & 0.0804 & 0.0803 & 0.1347 & 0.1031 & 0.1235 \\
 & $\bm{X}_{2}$ & 0.0705 & 0.0502 & 0.0485 & 0.0519 & --- & 0.0207 & 0.0400 & 0.0283 & 0.0353 \\
 & $\bm{X}_{3}$ & 0.0721 & 0.0501 & 0.0484 & 0.0519 & 0.0206 & --- & 0.0401 & 0.0280 & 0.0351 \\
 & $\bm{X}_{4}$ & 0.1867 & 0.1311 & 0.1301 & 0.1354 & 0.0646 & 0.0644 & --- & 0.0824 & 0.0998 \\
 & $\bm{X}_{5}$ & 0.0340 & 0.0381 & 0.0368 & 0.0395 & 0.0160 & 0.0159 & 0.0305 & --- & 0.0270 \\
 & $\bm{X}_{6}$ & 0.1649 & 0.1196 & 0.1156 & 0.1234 & 0.0560 & 0.0563 & 0.0990 & 0.0724 & --- \\  \hline
\multirow{8}{*}{PMM} & $\bm{X}_{12}$ & 0.2059 & --- & --- & --- & 0.0775 & 0.0776 & 0.1307 & 0.0996 & 0.1201 \\
 & $\bm{X}_{13}$ & 0.1970 & --- & --- & --- & 0.0744 & 0.0744 & 0.1295 & 0.0963 & 0.1159 \\
 & $\bm{X}_{14}$ & 0.2111 & --- & --- & --- & 0.0804 & 0.0803 & 0.1348 & 0.1031 & 0.1235 \\
 & $\bm{X}_{2}$ & 0.0704 & 0.0502 & 0.0485 & 0.0519 & --- & 0.0207 & 0.0400 & 0.0283 & 0.0353 \\
 & $\bm{X}_{3}$ & 0.0721 & 0.0501 & 0.0484 & 0.0519 & 0.0206 & --- & 0.0401 & 0.0280 & 0.0351 \\
 & $\bm{X}_{4}$ & 0.1860 & 0.1313 & 0.1303 & 0.1355 & 0.0647 & 0.0644 & --- & 0.0823 & 0.0999 \\
 & $\bm{X}_{5}$ & 0.0340 & 0.0381 & 0.0368 & 0.0395 & 0.0160 & 0.0159 & 0.0305 & --- & 0.0270 \\
 & $\bm{X}_{6}$ & 0.1649 & 0.1198 & 0.1157 & 0.1235 & 0.0560 & 0.0563 & 0.0991 & 0.0724 & --- \\  \hline
\multirow{8}{*}{CART} & $\bm{X}_{12}$ & 0.2056 & --- & --- & --- & 0.0774 & 0.0776 & 0.1311 & 0.1001 & 0.1205 \\
 & $\bm{X}_{13}$ & 0.1966 & --- & --- & --- & 0.0744 & 0.0745 & 0.1299 & 0.0967 & 0.1163 \\
 & $\bm{X}_{14}$ & 0.2125 & --- & --- & --- & 0.0805 & 0.0805 & 0.1355 & 0.1038 & 0.1243 \\
 & $\bm{X}_{2}$ & 0.0704 & 0.0503 & 0.0485 & 0.0520 & --- & 0.0207 & 0.0401 & 0.0283 & 0.0353 \\
 & $\bm{X}_{3}$ & 0.0719 & 0.0501 & 0.0484 & 0.0520 & 0.0205 & --- & 0.0402 & 0.0280 & 0.0351 \\
 & $\bm{X}_{4}$ & 0.1871 & 0.1317 & 0.1307 & 0.1361 & 0.0647 & 0.0645 & --- & 0.0827 & 0.1002 \\
 & $\bm{X}_{5}$ & 0.0340 & 0.0381 & 0.0368 & 0.0395 & 0.0160 & 0.0159 & 0.0304 & --- & 0.0270 \\
 & $\bm{X}_{6}$ & 0.1650 & 0.1202 & 0.1161 & 0.1242 & 0.0560 & 0.0564 & 0.0994 & 0.0727 & --- \\  \hline
\multirow{8}{*}{GERBIL} & $\bm{X}_{12}$ & 0.2047 & --- & --- & --- & 0.0774 & 0.0775 & 0.1305 & 0.0995 & 0.1199 \\
 & $\bm{X}_{13}$ & 0.1969 & --- & --- & --- & 0.0743 & 0.0744 & 0.1293 & 0.0962 & 0.1158 \\
 & $\bm{X}_{14}$ & 0.2107 & --- & --- & --- & 0.0804 & 0.0803 & 0.1348 & 0.1031 & 0.1236 \\
 & $\bm{X}_{2}$ & 0.0704 & 0.0502 & 0.0485 & 0.0519 & --- & 0.0207 & 0.0400 & 0.0283 & 0.0353 \\
 & $\bm{X}_{3}$ & 0.0721 & 0.0501 & 0.0484 & 0.0520 & 0.0206 & --- & 0.0401 & 0.0280 & 0.0351 \\
 & $\bm{X}_{4}$ & 0.1858 & 0.1311 & 0.1300 & 0.1355 & 0.0646 & 0.0644 & --- & 0.0822 & 0.0997 \\
 & $\bm{X}_{5}$ & 0.0340 & 0.0381 & 0.0368 & 0.0395 & 0.0160 & 0.0159 & 0.0305 & --- & 0.0270 \\
 & $\bm{X}_{6}$ & 0.1647 & 0.1196 & 0.1156 & 0.1235 & 0.0560 & 0.0563 & 0.0989 & 0.0724 & --- \\  \hline
\end{tabular}
}
\end{table}


\begin{table}[!ht]
\centering
{\small
\renewcommand{\tabcolsep}{.12cm}
\caption{Coverage rates and rMSE for the means of the simulated variables (where $\{\bm{X}_{11},\ldots,\bm{X}_{14}\}$ are binary indicators created from $\bm{X}_1$) across the 5,000 simulated datasets under a MAR missingness mechanism.  
} \label{MAR1}
\vspace{.12in}
\begin{tabular}{clccccccccc}
\hline \hline
& & $\bm{X}_{11}$ & $\bm{X}_{12}$ & $\bm{X}_{13}$ & $\bm{X}_{14}$ & $\bm{X}_{2}$ & $\bm{X}_{3}$ & $\bm{X}_{4}$ & $\bm{X}_{5}$ & $\bm{X}_{6}$ \\  \hline
\multirow{6}{*}{Coverage} & sbgcop & 0.8116 & 0.4750 & 0.7740 & 0.5042 & --- & 0.0436 & 0.5316 & 0.4704 & 0.9108 \\
 & jomo & 0.9456 & 0.9510 & 0.9504 & 0.9520 & --- & 0.9436 & 0.9390 & 0.9186 & 0.9468 \\
 & Logistic & 0.9266 & 0.9344 & 0.9276 & 0.9338 & --- & 0.9524 & 0.9496 & 0.9332 & 0.9458 \\
 & PMM & 0.9514 & 0.9358 & 0.9222 & 0.9486 & --- & 0.9508 & 0.9474 & 0.9530 & 0.9442 \\
 & CART & 0.9288 & 0.9350 & 0.9316 & 0.9380 & --- & 0.9256 & 0.9248 & 0.9324 & 0.9276 \\
 & GERBIL & 0.9602 & 0.9526 & 0.9622 & 0.9494 & --- & 0.9542 & 0.9466 & 0.9498 & 0.9444 \\  \hline
\multirow{6}{*}{rMSE} & sbgcop & 0.0174 & 0.0265 & 0.0193 & 0.0242 & --- & 0.1044 & 0.0282 & 0.0431 & 0.0152 \\
 & jomo & 0.0123 & 0.0119 & 0.0124 & 0.0115 & --- & 0.0301 & 0.0145 & 0.0240 & 0.0143 \\
 & Logistic & 0.0123 & 0.0119 & 0.0123 & 0.0115 & --- & 0.0290 & 0.0138 & 0.0202 & 0.0142 \\
 & PMM & 0.0124 & 0.0123 & 0.0128 & 0.0117 & --- & 0.0293 & 0.0140 & 0.0203 & 0.0143 \\
 & CART & 0.0124 & 0.0120 & 0.0125 & 0.0117 & --- & 0.0303 & 0.0144 & 0.0209 & 0.0145 \\
 & GERBIL & 0.0115 & 0.0117 & 0.0116 & 0.0116 & --- & 0.0291 & 0.0139 & 0.0203 & 0.0143 \\  \hline
\end{tabular}
}
\end{table}

\begin{table}[!ht]
\centering
{
\footnotesize
\renewcommand{\tabcolsep}{.12cm}
\caption{\footnotesize Coverage rates using six methods of imputation for the parameters of the fully-specified regression models of the simulated variables (where $\{\bm{X}_{12},\ldots,\bm{X}_{14}\}$ are binary indicators created from $\bm{X}_1$) across the 5,000 simulated datasets under a MAR missingness mechanism. Rows indicate the outcome variable and columns indicate predictors.  
} \label{MAR2}
\vspace{.12in}
\begin{tabular}{cccccccccccc}
\hline \hline
 & & Intercept & $\bm{X}_{12}$ & $\bm{X}_{13}$ & $\bm{X}_{14}$ & $\bm{X}_{2}$ & $\bm{X}_{3}$ & $\bm{X}_{4}$ & $\bm{X}_{5}$ & $\bm{X}_{6}$ \\  \hline
\multirow{8}{*}{sbgcop} & $\bm{X}_{12}$ & 0.9564 & --- & --- & --- & 0.9290 & 0.9562 & 0.9338 & 0.9546 & 0.9378 \\
 & $\bm{X}_{13}$ & 0.9592 & --- & --- & --- & 0.9576 & 0.9578 & 0.9416 & 0.9508 & 0.9462 \\
 & $\bm{X}_{14}$ & 0.9558 & --- & --- & --- & 0.9110 & 0.9534 & 0.9386 & 0.9622 & 0.9116 \\
 & $\bm{X}_{2}$ & 0.6638 & 0.9130 & 0.9542 & 0.8636 & --- & 0.8002 & 0.9222 & 0.7816 & 0.9370 \\
 & $\bm{X}_{3}$ & 0.3708 & 0.9550 & 0.9560 & 0.9468 & 0.7872 & --- & 0.9182 & 0.7700 & 0.8998 \\
 & $\bm{X}_{4}$ & 0.5378 & 0.9306 & 0.9406 & 0.9364 & 0.8474 & 0.8712 & --- & 0.6898 & 0.9326 \\
 & $\bm{X}_{5}$ & 0.9170 & 0.9524 & 0.9488 & 0.9586 & 0.9072 & 0.5824 & 0.6668 & --- & 0.8588 \\
 & $\bm{X}_{6}$ & 0.9276 & 0.9408 & 0.9460 & 0.9122 & 0.8798 & 0.9430 & 0.9330 & 0.8630 & --- \\  \hline
\multirow{8}{*}{jomo} & $\bm{X}_{12}$ & 0.9222 & --- & --- & --- & 0.9528 & 0.9466 & 0.9440 & 0.9078 & 0.9492 \\
 & $\bm{X}_{13}$ & 0.9418 & --- & --- & --- & 0.9518 & 0.9564 & 0.9504 & 0.9354 & 0.9512 \\
 & $\bm{X}_{14}$ & 0.7646 & --- & --- & --- & 0.9480 & 0.9382 & 0.9514 & 0.7366 & 0.9444 \\
 & $\bm{X}_{2}$ & 0.9542 & 0.9500 & 0.9504 & 0.9438 & --- & 0.9308 & 0.9352 & 0.9576 & 0.9540 \\
 & $\bm{X}_{3}$ & 0.8126 & 0.9462 & 0.9538 & 0.9336 & 0.9184 & --- & 0.9286 & 0.7306 & 0.9422 \\
 & $\bm{X}_{4}$ & 0.8358 & 0.9440 & 0.9510 & 0.9514 & 0.9386 & 0.9424 & --- & 0.7650 & 0.9406 \\
 & $\bm{X}_{5}$ & 0.7732 & 0.9272 & 0.9374 & 0.8260 & 0.9460 & 0.8618 & 0.8942 & --- & 0.9308 \\
 & $\bm{X}_{6}$ & 0.9238 & 0.9488 & 0.9514 & 0.9444 & 0.9532 & 0.9384 & 0.9416 & 0.9062 & --- \\  \hline
\multirow{8}{*}{Logistic} & $\bm{X}_{12}$ & 0.9288 & --- & --- & --- & 0.9300 & 0.9386 & 0.9408 & 0.9310 & 0.9368 \\
 & $\bm{X}_{13}$ & 0.9284 & --- & --- & --- & 0.9296 & 0.9438 & 0.9414 & 0.9292 & 0.9414 \\
 & $\bm{X}_{14}$ & 0.9252 & --- & --- & --- & 0.9334 & 0.9374 & 0.9426 & 0.9264 & 0.9416 \\
 & $\bm{X}_{2}$ & 0.9334 & 0.9284 & 0.9266 & 0.9370 & --- & 0.9436 & 0.9396 & 0.9280 & 0.9490 \\
 & $\bm{X}_{3}$ & 0.9488 & 0.9398 & 0.9432 & 0.9362 & 0.9508 & --- & 0.9484 & 0.9470 & 0.9440 \\
 & $\bm{X}_{4}$ & 0.9498 & 0.9420 & 0.9424 & 0.9440 & 0.9406 & 0.9520 & --- & 0.9484 & 0.9456 \\
 & $\bm{X}_{5}$ & 0.9316 & 0.9272 & 0.9288 & 0.9264 & 0.9278 & 0.9410 & 0.9486 & --- & 0.9394 \\
 & $\bm{X}_{6}$ & 0.9388 & 0.9374 & 0.9404 & 0.9400 & 0.9478 & 0.9452 & 0.9460 & 0.9394 & --- \\  \hline
\multirow{8}{*}{PMM} & $\bm{X}_{12}$ & 0.9658 & --- & --- & --- & 0.9828 & 0.9720 & 0.9836 & 0.9764 & 0.9182 \\
 & $\bm{X}_{13}$ & 0.9778 & --- & --- & --- & 0.9816 & 0.9834 & 0.9862 & 0.9790 & 0.9472 \\
 & $\bm{X}_{14}$ & 0.9350 & --- & --- & --- & 0.9494 & 0.9310 & 0.9360 & 0.9392 & 0.9088 \\
 & $\bm{X}_{2}$ & 0.9484 & 0.9822 & 0.9790 & 0.9492 & --- & 0.9432 & 0.9430 & 0.9434 & 0.9506 \\
 & $\bm{X}_{3}$ & 0.9590 & 0.9746 & 0.9832 & 0.9324 & 0.9496 & --- & 0.9470 & 0.9510 & 0.9410 \\
 & $\bm{X}_{4}$ & 0.9528 & 0.9840 & 0.9864 & 0.9374 & 0.9442 & 0.9510 & --- & 0.9514 & 0.9432 \\
 & $\bm{X}_{5}$ & 0.9524 & 0.9754 & 0.9772 & 0.9384 & 0.9424 & 0.9548 & 0.9492 & --- & 0.9456 \\
 & $\bm{X}_{6}$ & 0.9518 & 0.9150 & 0.9468 & 0.9084 & 0.9520 & 0.9370 & 0.9450 & 0.9492 & --- \\  \hline
\multirow{8}{*}{CART} & $\bm{X}_{12}$ & 0.9020 & --- & --- & --- & 0.9096 & 0.9550 & 0.9704 & 0.9040 & 0.9666 \\
 & $\bm{X}_{13}$ & 0.9772 & --- & --- & --- & 0.9398 & 0.9734 & 0.9834 & 0.9666 & 0.8698 \\
 & $\bm{X}_{14}$ & 0.8478 & --- & --- & --- & 0.9096 & 0.9526 & 0.9616 & 0.8180 & 0.9172 \\
 & $\bm{X}_{2}$ & 0.8984 & 0.8994 & 0.9374 & 0.9006 & --- & 0.9010 & 0.9096 & 0.9150 & 0.9266 \\
 & $\bm{X}_{3}$ & 0.9080 & 0.9562 & 0.9708 & 0.9510 & 0.8842 & --- & 0.8608 & 0.8982 & 0.8928 \\
 & $\bm{X}_{4}$ & 0.8286 & 0.9700 & 0.9860 & 0.9610 & 0.9112 & 0.8762 & --- & 0.7896 & 0.9134 \\
 & $\bm{X}_{5}$ & 0.6636 & 0.9132 & 0.9684 & 0.8470 & 0.8908 & 0.9016 & 0.8156 & --- & 0.8386 \\
 & $\bm{X}_{6}$ & 0.9300 & 0.9670 & 0.8652 & 0.9176 & 0.9252 & 0.8866 & 0.9122 & 0.8170 & --- \\  \hline
\multirow{8}{*}{GERBIL} & $\bm{X}_{12}$ & 0.9548 & --- & --- & --- & 0.9512 & 0.9478 & 0.9502 & 0.9494 & 0.9488 \\
 & $\bm{X}_{13}$ & 0.9602 & --- & --- & --- & 0.9506 & 0.9500 & 0.9538 & 0.9530 & 0.9534 \\
 & $\bm{X}_{14}$ & 0.9454 & --- & --- & --- & 0.9544 & 0.9400 & 0.9524 & 0.9462 & 0.9516 \\
 & $\bm{X}_{2}$ & 0.9474 & 0.9486 & 0.9516 & 0.9554 & --- & 0.9476 & 0.9366 & 0.9454 & 0.9510 \\
 & $\bm{X}_{3}$ & 0.9524 & 0.9476 & 0.9488 & 0.9430 & 0.9506 & --- & 0.9494 & 0.9518 & 0.9452 \\
 & $\bm{X}_{4}$ & 0.9524 & 0.9504 & 0.9540 & 0.9528 & 0.9390 & 0.9496 & --- & 0.9522 & 0.9434 \\
 & $\bm{X}_{5}$ & 0.9480 & 0.9470 & 0.9516 & 0.9502 & 0.9428 & 0.9484 & 0.9520 & --- & 0.9460 \\
 & $\bm{X}_{6}$ & 0.9484 & 0.9484 & 0.9538 & 0.9524 & 0.9506 & 0.9464 & 0.9438 & 0.9476 & --- \\  \hline
\end{tabular}
}
\end{table}

\begin{table}[!ht]
\centering
{
\footnotesize
\renewcommand{\tabcolsep}{.12cm}
\caption{\footnotesize rMSE using six methods of imputation for the parameters of the fully-specified regression models of the simulated variables (where $\{\bm{X}_{12},\ldots,\bm{X}_{14}\}$ are binary indicators created from $\bm{X}_1$) across the 5,000 simulated datasets under a MAR missingness mechanism. Rows indicate the outcome variable and columns indicate predictors.  
} \label{MAR3}
\vspace{.12in}
\begin{tabular}{cccccccccccc}
\hline \hline
 & & Intercept & $\bm{X}_{12}$ & $\bm{X}_{13}$ & $\bm{X}_{14}$ & $\bm{X}_{2}$ & $\bm{X}_{3}$ & $\bm{X}_{4}$ & $\bm{X}_{5}$ & $\bm{X}_{6}$ \\  \hline
\multirow{8}{*}{sbgcop} & $\bm{X}_{12}$ & 0.4153 & --- & --- & --- & 0.1171 & 0.1266 & 0.2741 & 0.1747 & 0.2397 \\
 & $\bm{X}_{13}$ & 0.3904 & --- & --- & --- & 0.1047 & 0.1290 & 0.2632 & 0.1741 & 0.2240 \\
 & $\bm{X}_{14}$ & 0.4234 & --- & --- & --- & 0.1317 & 0.1298 & 0.2859 & 0.1752 & 0.2731 \\
 & $\bm{X}_{2}$ & 0.1798 & 0.0824 & 0.0705 & 0.0975 & --- & 0.0447 & 0.0676 & 0.0621 & 0.0548 \\
 & $\bm{X}_{3}$ & 0.3035 & 0.0840 & 0.0863 & 0.0882 & 0.0458 & --- & 0.0820 & 0.0753 & 0.0721 \\
 & $\bm{X}_{4}$ & 0.8349 & 0.2770 & 0.2664 & 0.2897 & 0.1257 & 0.1443 & --- & 0.2669 & 0.2034 \\
 & $\bm{X}_{5}$ & 0.0608 & 0.0614 & 0.0610 & 0.0615 & 0.0253 & 0.0503 & 0.0901 & --- & 0.0586 \\
 & $\bm{X}_{6}$ & 0.3581 & 0.2384 & 0.2231 & 0.2726 & 0.0983 & 0.1016 & 0.2000 & 0.1715 & --- \\  \hline
\multirow{8}{*}{jomo} & $\bm{X}_{12}$ & 0.4721 & --- & --- & --- & 0.1150 & 0.1339 & 0.2506 & 0.1908 & 0.2203 \\
 & $\bm{X}_{13}$ & 0.4238 & --- & --- & --- & 0.1118 & 0.1307 & 0.2408 & 0.1697 & 0.2082 \\
 & $\bm{X}_{14}$ & 0.6952 & --- & --- & --- & 0.1216 & 0.1420 & 0.2559 & 0.2754 & 0.2291 \\
 & $\bm{X}_{2}$ & 0.0985 & 0.0724 & 0.0711 & 0.0763 & --- & 0.0311 & 0.0629 & 0.0378 & 0.0500 \\
 & $\bm{X}_{3}$ & 0.1821 & 0.0849 & 0.0840 & 0.0904 & 0.0349 & --- & 0.0742 & 0.0745 & 0.0605 \\
 & $\bm{X}_{4}$ & 0.5389 & 0.2518 & 0.2414 & 0.2582 & 0.1034 & 0.1201 & --- & 0.2216 & 0.1820 \\
 & $\bm{X}_{5}$ & 0.0915 & 0.0686 & 0.0644 & 0.0910 & 0.0244 & 0.0346 & 0.0654 & --- & 0.0479 \\
 & $\bm{X}_{6}$ & 0.3504 & 0.2197 & 0.2078 & 0.2288 & 0.0827 & 0.1002 & 0.1794 & 0.1365 & --- \\  \hline
\multirow{8}{*}{Logistic} & $\bm{X}_{12}$ & 0.4501 & --- & --- & --- & 0.1161 & 0.1397 & 0.2525 & 0.1810 & 0.2241 \\
 & $\bm{X}_{13}$ & 0.4406 & --- & --- & --- & 0.1152 & 0.1399 & 0.2454 & 0.1800 & 0.2127 \\
 & $\bm{X}_{14}$ & 0.4724 & --- & --- & --- & 0.1197 & 0.1444 & 0.2619 & 0.1867 & 0.2325 \\
 & $\bm{X}_{2}$ & 0.1032 & 0.0723 & 0.0724 & 0.0740 & --- & 0.0294 & 0.0616 & 0.0409 & 0.0498 \\
 & $\bm{X}_{3}$ & 0.1195 & 0.0866 & 0.0876 & 0.0892 & 0.0308 & --- & 0.0707 & 0.0439 & 0.0601 \\
 & $\bm{X}_{4}$ & 0.3917 & 0.2533 & 0.2461 & 0.2632 & 0.1027 & 0.1184 & --- & 0.1457 & 0.1835 \\
 & $\bm{X}_{5}$ & 0.0569 & 0.0641 & 0.0634 & 0.0659 & 0.0236 & 0.0246 & 0.0507 & --- & 0.0458 \\
 & $\bm{X}_{6}$ & 0.3393 & 0.2238 & 0.2127 & 0.2324 & 0.0832 & 0.1011 & 0.1822 & 0.1312 & --- \\  \hline
\multirow{8}{*}{PMM} & $\bm{X}_{12}$ & 0.4167 & --- & --- & --- & 0.0960 & 0.1216 & 0.2054 & 0.1560 & 0.2551 \\
 & $\bm{X}_{13}$ & 0.3612 & --- & --- & --- & 0.0901 & 0.1099 & 0.1829 & 0.1450 & 0.2112 \\
 & $\bm{X}_{14}$ & 0.5096 & --- & --- & --- & 0.1264 & 0.1532 & 0.2866 & 0.1990 & 0.2802 \\
 & $\bm{X}_{2}$ & 0.1034 & 0.0600 & 0.0568 & 0.0784 & --- & 0.0300 & 0.0617 & 0.0416 & 0.0500 \\
 & $\bm{X}_{3}$ & 0.1189 & 0.0759 & 0.0687 & 0.0947 & 0.0314 & --- & 0.0698 & 0.0448 & 0.0618 \\
 & $\bm{X}_{4}$ & 0.4052 & 0.2054 & 0.1826 & 0.2876 & 0.1018 & 0.1167 & --- & 0.1523 & 0.1820 \\
 & $\bm{X}_{5}$ & 0.0562 & 0.0551 & 0.0509 & 0.0696 & 0.0237 & 0.0249 & 0.0524 & --- & 0.0459 \\
 & $\bm{X}_{6}$ & 0.3363 & 0.2562 & 0.2111 & 0.2805 & 0.0835 & 0.1029 & 0.1805 & 0.1329 & --- \\  \hline
\multirow{8}{*}{CART} & $\bm{X}_{12}$ & 0.4748 & --- & --- & --- & 0.1282 & 0.1206 & 0.1995 & 0.1868 & 0.1795 \\
 & $\bm{X}_{13}$ & 0.3401 & --- & --- & --- & 0.1115 & 0.1151 & 0.1715 & 0.1460 & 0.2485 \\
 & $\bm{X}_{14}$ & 0.5646 & --- & --- & --- & 0.1322 & 0.1265 & 0.2215 & 0.2276 & 0.2312 \\
 & $\bm{X}_{2}$ & 0.1206 & 0.0810 & 0.0704 & 0.0832 & --- & 0.0343 & 0.0663 & 0.0449 & 0.0511 \\
 & $\bm{X}_{3}$ & 0.1444 & 0.0756 & 0.0730 & 0.0787 & 0.0379 & --- & 0.0835 & 0.0547 & 0.0680 \\
 & $\bm{X}_{4}$ & 0.5289 & 0.2007 & 0.1699 & 0.2252 & 0.1098 & 0.1374 & --- & 0.2103 & 0.1893 \\
 & $\bm{X}_{5}$ & 0.0922 & 0.0657 & 0.0523 & 0.0788 & 0.0287 & 0.0309 & 0.0723 & --- & 0.0559 \\
 & $\bm{X}_{6}$ & 0.3335 & 0.1790 & 0.2487 & 0.2305 & 0.0846 & 0.1113 & 0.1873 & 0.1600 & --- \\  \hline
\multirow{8}{*}{GERBIL} & $\bm{X}_{12}$ & 0.4441 & --- & --- & --- & 0.1152 & 0.1391 & 0.2535 & 0.1803 & 0.2258 \\
 & $\bm{X}_{13}$ & 0.4158 & --- & --- & --- & 0.1117 & 0.1376 & 0.2471 & 0.1741 & 0.2124 \\
 & $\bm{X}_{14}$ & 0.4707 & --- & --- & --- & 0.1195 & 0.1467 & 0.2607 & 0.1877 & 0.2331 \\
 & $\bm{X}_{2}$ & 0.1031 & 0.0715 & 0.0697 & 0.0738 & --- & 0.0294 & 0.0619 & 0.0412 & 0.0500 \\
 & $\bm{X}_{3}$ & 0.1188 & 0.0853 & 0.0854 & 0.0899 & 0.0308 & --- & 0.0709 & 0.0439 & 0.0605 \\
 & $\bm{X}_{4}$ & 0.4013 & 0.2565 & 0.2505 & 0.2639 & 0.1039 & 0.1213 & --- & 0.1497 & 0.1878 \\
 & $\bm{X}_{5}$ & 0.0564 & 0.0633 & 0.0608 & 0.0657 & 0.0236 & 0.0247 & 0.0512 & --- & 0.0462 \\
 & $\bm{X}_{6}$ & 0.3437 & 0.2248 & 0.2116 & 0.2326 & 0.0838 & 0.1025 & 0.1840 & 0.1333 & --- \\  \hline
\end{tabular}
}
\end{table}

\begin{table}[!ht]
\centering
{
\scriptsize
\renewcommand{\tabcolsep}{.12cm}
\caption{\footnotesize rMSE using six methods of imputation for elements of the variance-covariance matrix calculated with the simulated variables (where $\{\bm{X}_{11},\ldots,\bm{X}_{14}\}$ are binary indicators created from $\bm{X}_1$) across the 5,000 simulated datasets under a MAR missingness mechanism. 
} \label{MAR4}
\vspace{.12in}
\begin{tabular}{ccccccccccccc}
\hline \hline
 & & $\bm{X}_{11}$ & $\bm{X}_{12}$ & $\bm{X}_{13}$ & $\bm{X}_{14}$ & $\bm{X}_{2}$ & $\bm{X}_{3}$ & $\bm{X}_{4}$ & $\bm{X}_{5}$ & $\bm{X}_{6}$ \\  \hline
\multirow{9}{*}{sbgcop} & $\bm{X}_{11}$ & 0.0088 & --- & --- & --- & --- & --- & --- & --- & --- \\
 & $\bm{X}_{12}$ & 0.0068 & 0.0129 & --- & --- & --- & --- & --- & --- & --- \\
 & $\bm{X}_{13}$ & 0.0049 & 0.0071 & 0.0094 & --- & --- & --- & --- & --- & --- \\
 & $\bm{X}_{14}$ & 0.0070 & 0.0048 & 0.0068 & 0.0121 & --- & --- & --- & --- & --- \\
 & $\bm{X}_{2}$ & 0.0136 & 0.0126 & 0.0126 & 0.0119 & --- & --- & --- & --- & --- \\
 & $\bm{X}_{3}$ & 0.0192 & 0.0148 & 0.0145 & 0.0166 & 0.0895 & 0.0765 & --- & --- & --- \\
 & $\bm{X}_{4}$ & 0.0079 & 0.0087 & 0.0075 & 0.0093 & 0.0223 & 0.0368 & 0.0011 & --- & --- \\
 & $\bm{X}_{5}$ & 0.0115 & 0.0098 & 0.0097 & 0.0114 & 0.0264 & 0.0728 & 0.0108 & 0.0251 & --- \\
 & $\bm{X}_{6}$ & 0.0076 & 0.0080 & 0.0076 & 0.0082 & 0.0154 & 0.0203 & 0.0087 & 0.0147 & 0.0004 \\  \hline
\multirow{9}{*}{jomo} & $\bm{X}_{11}$ & 0.0062 & --- & --- & --- & --- & --- & --- & --- & --- \\
 & $\bm{X}_{12}$ & 0.0035 & 0.0059 & --- & --- & --- & --- & --- & --- & --- \\
 & $\bm{X}_{13}$ & 0.0034 & 0.0035 & 0.0062 & --- & --- & --- & --- & --- & --- \\
 & $\bm{X}_{14}$ & 0.0035 & 0.0034 & 0.0035 & 0.0057 & --- & --- & --- & --- & --- \\
 & $\bm{X}_{2}$ & 0.0141 & 0.0123 & 0.0131 & 0.0128 & --- & --- & --- & --- & --- \\
 & $\bm{X}_{3}$ & 0.0166 & 0.0139 & 0.0145 & 0.0149 & 0.0365 & 0.0466 & --- & --- & --- \\
 & $\bm{X}_{4}$ & 0.0074 & 0.0074 & 0.0073 & 0.0080 & 0.0142 & 0.0175 & 0.0003 & --- & --- \\
 & $\bm{X}_{5}$ & 0.0171 & 0.0099 & 0.0114 & 0.0187 & 0.0384 & 0.0550 & 0.0239 & 0.0233 & --- \\
 & $\bm{X}_{6}$ & 0.0072 & 0.0073 & 0.0072 & 0.0071 & 0.0152 & 0.0175 & 0.0082 & 0.0119 & 0.0003 \\  \hline
\multirow{9}{*}{Logistic} & $\bm{X}_{11}$ & 0.0061 & --- & --- & --- & --- & --- & --- & --- & --- \\
 & $\bm{X}_{12}$ & 0.0035 & 0.0059 & --- & --- & --- & --- & --- & --- & --- \\
 & $\bm{X}_{13}$ & 0.0034 & 0.0035 & 0.0062 & --- & --- & --- & --- & --- & --- \\
 & $\bm{X}_{14}$ & 0.0035 & 0.0034 & 0.0035 & 0.0058 & --- & --- & --- & --- & --- \\
 & $\bm{X}_{2}$ & 0.0138 & 0.0122 & 0.0133 & 0.0118 & --- & --- & --- & --- & --- \\
 & $\bm{X}_{3}$ & 0.0156 & 0.0137 & 0.0148 & 0.0134 & 0.0351 & 0.0459 & --- & --- & --- \\
 & $\bm{X}_{4}$ & 0.0069 & 0.0073 & 0.0072 & 0.0070 & 0.0131 & 0.0154 & 0.0003 & --- & --- \\
 & $\bm{X}_{5}$ & 0.0110 & 0.0099 & 0.0104 & 0.0097 & 0.0237 & 0.0260 & 0.0111 & 0.0213 & --- \\
 & $\bm{X}_{6}$ & 0.0072 & 0.0073 & 0.0073 & 0.0072 & 0.0152 & 0.0175 & 0.0082 & 0.0125 & 0.0003 \\  \hline
\multirow{9}{*}{PMM} & $\bm{X}_{11}$ & 0.0062 & --- & --- & --- & --- & --- & --- & --- & --- \\
 & $\bm{X}_{12}$ & 0.0036 & 0.0061 & --- & --- & --- & --- & --- & --- & --- \\
 & $\bm{X}_{13}$ & 0.0036 & 0.0035 & 0.0065 & --- & --- & --- & --- & --- & --- \\
 & $\bm{X}_{14}$ & 0.0035 & 0.0036 & 0.0036 & 0.0058 & --- & --- & --- & --- & --- \\
 & $\bm{X}_{2}$ & 0.0136 & 0.0129 & 0.0141 & 0.0124 & --- & --- & --- & --- & --- \\
 & $\bm{X}_{3}$ & 0.0153 & 0.0138 & 0.0145 & 0.0139 & 0.0365 & 0.0490 & --- & --- & --- \\
 & $\bm{X}_{4}$ & 0.0067 & 0.0063 & 0.0062 & 0.0071 & 0.0134 & 0.0160 & 0.0003 & --- & --- \\
 & $\bm{X}_{5}$ & 0.0110 & 0.0095 & 0.0101 & 0.0100 & 0.0244 & 0.0277 & 0.0113 & 0.0217 & --- \\
 & $\bm{X}_{6}$ & 0.0069 & 0.0087 & 0.0109 & 0.0078 & 0.0152 & 0.0176 & 0.0081 & 0.0126 & 0.0003 \\  \hline
\multirow{9}{*}{CART} & $\bm{X}_{11}$ & 0.0062 & --- & --- & --- & --- & --- & --- & --- & --- \\
 & $\bm{X}_{12}$ & 0.0036 & 0.0060 & --- & --- & --- & --- & --- & --- & --- \\
 & $\bm{X}_{13}$ & 0.0034 & 0.0035 & 0.0062 & --- & --- & --- & --- & --- & --- \\
 & $\bm{X}_{14}$ & 0.0035 & 0.0034 & 0.0036 & 0.0058 & --- & --- & --- & --- & --- \\
 & $\bm{X}_{2}$ & 0.0142 & 0.0124 & 0.0137 & 0.0120 & --- & --- & --- & --- & --- \\
 & $\bm{X}_{3}$ & 0.0173 & 0.0147 & 0.0149 & 0.0148 & 0.0399 & 0.0518 & --- & --- & --- \\
 & $\bm{X}_{4}$ & 0.0074 & 0.0067 & 0.0064 & 0.0074 & 0.0147 & 0.0175 & 0.0003 & --- & --- \\
 & $\bm{X}_{5}$ & 0.0141 & 0.0101 & 0.0088 & 0.0118 & 0.0266 & 0.0315 & 0.0168 & 0.0223 & --- \\
 & $\bm{X}_{6}$ & 0.0081 & 0.0067 & 0.0083 & 0.0071 & 0.0160 & 0.0194 & 0.0083 & 0.0131 & 0.0003 \\  \hline
\multirow{9}{*}{GERBIL} & $\bm{X}_{11}$ & 0.0058 & --- & --- & --- & --- & --- & --- & --- & --- \\
 & $\bm{X}_{12}$ & 0.0034 & 0.0059 & --- & --- & --- & --- & --- & --- & --- \\
 & $\bm{X}_{13}$ & 0.0033 & 0.0033 & 0.0058 & --- & --- & --- & --- & --- & --- \\
 & $\bm{X}_{14}$ & 0.0033 & 0.0033 & 0.0034 & 0.0058 & --- & --- & --- & --- & --- \\
 & $\bm{X}_{2}$ & 0.0130 & 0.0122 & 0.0128 & 0.0118 & --- & --- & --- & --- & --- \\
 & $\bm{X}_{3}$ & 0.0148 & 0.0138 & 0.0145 & 0.0136 & 0.0351 & 0.0459 & --- & --- & --- \\
 & $\bm{X}_{4}$ & 0.0069 & 0.0073 & 0.0071 & 0.0069 & 0.0131 & 0.0153 & 0.0003 & --- & --- \\
 & $\bm{X}_{5}$ & 0.0104 & 0.0100 & 0.0101 & 0.0098 & 0.0236 & 0.0257 & 0.0109 & 0.0213 & --- \\
 & $\bm{X}_{6}$ & 0.0072 & 0.0074 & 0.0073 & 0.0072 & 0.0151 & 0.0176 & 0.0082 & 0.0125 & 0.0003 \\  \hline
\end{tabular}
}
\end{table}

\begin{table}[!ht]
\centering
{
\footnotesize
\renewcommand{\tabcolsep}{.12cm}
\caption{\footnotesize rMSE using six methods of imputation for the squared standard errors of the parameters of the fully-specified regression models with the simulated variables (where $\{\bm{X}_{12},\ldots,\bm{X}_{14}\}$ are binary indicators created from $\bm{X}_1$) across the 5,000 simulated datasets under a MAR missingness mechanism. Rows indicate the outcome variable and columns indicate predictors.  
} \label{MAR5}
\vspace{.12in}
\begin{tabular}{cccccccccccc}
\hline \hline
 & & Intercept & $\bm{X}_{12}$ & $\bm{X}_{13}$ & $\bm{X}_{14}$ & $\bm{X}_{2}$ & $\bm{X}_{3}$ & $\bm{X}_{4}$ & $\bm{X}_{5}$ & $\bm{X}_{6}$ \\  \hline
\multirow{8}{*}{sbgcop} & $\bm{X}_{12}$ & 0.2071 & --- & --- & --- & 0.0778 & 0.0779 & 0.1306 & 0.0995 & 0.1196 \\
 & $\bm{X}_{13}$ & 0.1983 & --- & --- & --- & 0.0747 & 0.0748 & 0.1288 & 0.0961 & 0.1153 \\
 & $\bm{X}_{14}$ & 0.2113 & --- & --- & --- & 0.0807 & 0.0806 & 0.1346 & 0.1028 & 0.1228 \\
 & $\bm{X}_{2}$ & 0.0703 & 0.0500 & 0.0483 & 0.0516 & --- & 0.0207 & 0.0400 & 0.0282 & 0.0352 \\
 & $\bm{X}_{3}$ & 0.0715 & 0.0499 & 0.0482 & 0.0517 & 0.0205 & --- & 0.0401 & 0.0279 & 0.0350 \\
 & $\bm{X}_{4}$ & 0.1844 & 0.1313 & 0.1297 & 0.1354 & 0.0651 & 0.0649 & --- & 0.0823 & 0.1000 \\
 & $\bm{X}_{5}$ & 0.0340 & 0.0381 & 0.0368 & 0.0395 & 0.0160 & 0.0159 & 0.0305 & --- & 0.0270 \\
 & $\bm{X}_{6}$ & 0.1647 & 0.1194 & 0.1151 & 0.1227 & 0.0562 & 0.0565 & 0.0991 & 0.0723 & --- \\  \hline
\multirow{8}{*}{jomo} & $\bm{X}_{12}$ & 0.2081 & --- & --- & --- & 0.0775 & 0.0777 & 0.1310 & 0.1007 & 0.1201 \\
 & $\bm{X}_{13}$ & 0.1993 & --- & --- & --- & 0.0744 & 0.0746 & 0.1297 & 0.0972 & 0.1159 \\
 & $\bm{X}_{14}$ & 0.2173 & --- & --- & --- & 0.0806 & 0.0806 & 0.1357 & 0.1045 & 0.1240 \\
 & $\bm{X}_{2}$ & 0.0705 & 0.0502 & 0.0485 & 0.0519 & --- & 0.0207 & 0.0401 & 0.0283 & 0.0353 \\
 & $\bm{X}_{3}$ & 0.0719 & 0.0500 & 0.0484 & 0.0519 & 0.0205 & --- & 0.0402 & 0.0280 & 0.0351 \\
 & $\bm{X}_{4}$ & 0.1909 & 0.1318 & 0.1307 & 0.1365 & 0.0648 & 0.0646 & --- & 0.0833 & 0.1002 \\
 & $\bm{X}_{5}$ & 0.0339 & 0.0380 & 0.0366 & 0.0394 & 0.0160 & 0.0159 & 0.0304 & --- & 0.0269 \\
 & $\bm{X}_{6}$ & 0.1659 & 0.1198 & 0.1156 & 0.1240 & 0.0560 & 0.0564 & 0.0994 & 0.0730 & --- \\  \hline
\multirow{8}{*}{Logistic} & $\bm{X}_{12}$ & 0.2048 & --- & --- & --- & 0.0774 & 0.0775 & 0.1305 & 0.0996 & 0.1199 \\
 & $\bm{X}_{13}$ & 0.1969 & --- & --- & --- & 0.0743 & 0.0744 & 0.1293 & 0.0963 & 0.1158 \\
 & $\bm{X}_{14}$ & 0.2106 & --- & --- & --- & 0.0804 & 0.0803 & 0.1348 & 0.1032 & 0.1235 \\
 & $\bm{X}_{2}$ & 0.0705 & 0.0502 & 0.0485 & 0.0519 & --- & 0.0207 & 0.0401 & 0.0283 & 0.0353 \\
 & $\bm{X}_{3}$ & 0.0721 & 0.0501 & 0.0484 & 0.0519 & 0.0205 & --- & 0.0402 & 0.0280 & 0.0351 \\
 & $\bm{X}_{4}$ & 0.1868 & 0.1312 & 0.1302 & 0.1355 & 0.0647 & 0.0645 & --- & 0.0824 & 0.0998 \\
 & $\bm{X}_{5}$ & 0.0340 & 0.0381 & 0.0368 & 0.0395 & 0.0160 & 0.0159 & 0.0305 & --- & 0.0270 \\
 & $\bm{X}_{6}$ & 0.1649 & 0.1196 & 0.1156 & 0.1234 & 0.0560 & 0.0563 & 0.0990 & 0.0725 & --- \\  \hline
\multirow{8}{*}{PMM} & $\bm{X}_{12}$ & 0.2064 & --- & --- & --- & 0.0775 & 0.0776 & 0.1308 & 0.0996 & 0.1202 \\
 & $\bm{X}_{13}$ & 0.1964 & --- & --- & --- & 0.0743 & 0.0744 & 0.1293 & 0.0962 & 0.1158 \\
 & $\bm{X}_{14}$ & 0.2114 & --- & --- & --- & 0.0804 & 0.0803 & 0.1348 & 0.1031 & 0.1236 \\
 & $\bm{X}_{2}$ & 0.0704 & 0.0502 & 0.0485 & 0.0519 & --- & 0.0207 & 0.0401 & 0.0283 & 0.0353 \\
 & $\bm{X}_{3}$ & 0.0721 & 0.0501 & 0.0484 & 0.0519 & 0.0205 & --- & 0.0402 & 0.0280 & 0.0351 \\
 & $\bm{X}_{4}$ & 0.1860 & 0.1313 & 0.1301 & 0.1355 & 0.0647 & 0.0645 & --- & 0.0823 & 0.0999 \\
 & $\bm{X}_{5}$ & 0.0340 & 0.0381 & 0.0368 & 0.0395 & 0.0160 & 0.0159 & 0.0305 & --- & 0.0270 \\
 & $\bm{X}_{6}$ & 0.1650 & 0.1200 & 0.1156 & 0.1235 & 0.0560 & 0.0563 & 0.0991 & 0.0724 & --- \\  \hline
\multirow{8}{*}{CART} & $\bm{X}_{12}$ & 0.2055 & --- & --- & --- & 0.0775 & 0.0777 & 0.1311 & 0.1001 & 0.1206 \\
 & $\bm{X}_{13}$ & 0.1963 & --- & --- & --- & 0.0744 & 0.0745 & 0.1299 & 0.0967 & 0.1163 \\
 & $\bm{X}_{14}$ & 0.2125 & --- & --- & --- & 0.0805 & 0.0805 & 0.1357 & 0.1038 & 0.1244 \\
 & $\bm{X}_{2}$ & 0.0703 & 0.0502 & 0.0485 & 0.0520 & --- & 0.0207 & 0.0401 & 0.0283 & 0.0353 \\
 & $\bm{X}_{3}$ & 0.0719 & 0.0501 & 0.0484 & 0.0520 & 0.0205 & --- & 0.0402 & 0.0280 & 0.0351 \\
 & $\bm{X}_{4}$ & 0.1874 & 0.1318 & 0.1310 & 0.1364 & 0.0648 & 0.0646 & --- & 0.0828 & 0.1004 \\
 & $\bm{X}_{5}$ & 0.0340 & 0.0381 & 0.0367 & 0.0395 & 0.0160 & 0.0159 & 0.0305 & --- & 0.0270 \\
 & $\bm{X}_{6}$ & 0.1650 & 0.1203 & 0.1161 & 0.1243 & 0.0561 & 0.0564 & 0.0995 & 0.0728 & --- \\  \hline
\multirow{8}{*}{GERBIL} & $\bm{X}_{12}$ & 0.2044 & --- & --- & --- & 0.0774 & 0.0774 & 0.1304 & 0.0995 & 0.1198 \\
 & $\bm{X}_{13}$ & 0.1968 & --- & --- & --- & 0.0743 & 0.0743 & 0.1292 & 0.0962 & 0.1157 \\
 & $\bm{X}_{14}$ & 0.2102 & --- & --- & --- & 0.0803 & 0.0802 & 0.1347 & 0.1030 & 0.1235 \\
 & $\bm{X}_{2}$ & 0.0705 & 0.0502 & 0.0485 & 0.0519 & --- & 0.0207 & 0.0401 & 0.0283 & 0.0353 \\
 & $\bm{X}_{3}$ & 0.0721 & 0.0501 & 0.0484 & 0.0519 & 0.0205 & --- & 0.0402 & 0.0280 & 0.0351 \\
 & $\bm{X}_{4}$ & 0.1857 & 0.1310 & 0.1299 & 0.1353 & 0.0646 & 0.0644 & --- & 0.0822 & 0.0997 \\
 & $\bm{X}_{5}$ & 0.0340 & 0.0381 & 0.0368 & 0.0395 & 0.0160 & 0.0159 & 0.0305 & --- & 0.0270 \\
 & $\bm{X}_{6}$ & 0.1647 & 0.1196 & 0.1155 & 0.1234 & 0.0560 & 0.0563 & 0.0990 & 0.0724 & --- \\  \hline
\end{tabular}
}
\end{table}


\begin{table}[!ht]
\centering
{\small
\renewcommand{\tabcolsep}{.12cm}
\caption{Coverage rates and rMSE for the means of the simulated variables (where $\{\bm{X}_{11},\ldots,\bm{X}_{14}\}$ are binary indicators created from $\bm{X}_1$) across the 5,000 simulated datasets under a NMAR missingness mechanism.  
} \label{NMAR1}
\vspace{.12in}
\begin{tabular}{clccccccccc}
\hline \hline
& & $\bm{X}_{11}$ & $\bm{X}_{12}$ & $\bm{X}_{13}$ & $\bm{X}_{14}$ & $\bm{X}_{2}$ & $\bm{X}_{3}$ & $\bm{X}_{4}$ & $\bm{X}_{5}$ & $\bm{X}_{6}$ \\  \hline
\multirow{6}{*}{Coverage} & sbgcop & 0.2358 & 0.8952 & 0.7146 & 0.7562 & --- & 0.0000 & 0.0014 & 0.6686 & 0.3302 \\
 & jomo & 0.3278 & 0.9050 & 0.7870 & 0.6018 & --- & 0.0000 & 0.0000 & 0.7336 & 0.0918 \\
 & Logistic & 0.3212 & 0.8976 & 0.7794 & 0.5856 & --- & 0.0000 & 0.0000 & 0.7486 & 0.1040 \\
 & PMM & 0.3286 & 0.8990 & 0.7930 & 0.5878 & --- & 0.0000 & 0.0000 & 0.7506 & 0.1070 \\
 & CART & 0.3106 & 0.8990 & 0.7802 & 0.5782 & --- & 0.0000 & 0.0000 & 0.7330 & 0.0738 \\
 & GERBIL & 0.3264 & 0.9030 & 0.7916 & 0.5902 & --- & 0.0000 & 0.0000 & 0.7522 & 0.1048 \\  \hline
\multirow{6}{*}{rMSE} & sbgcop & 0.0291 & 0.0122 & 0.0179 & 0.0167 & --- & 0.2916 & 0.0717 & 0.0313 & 0.0371 \\
 & jomo & 0.0265 & 0.0121 & 0.0164 & 0.0211 & --- & 0.2217 & 0.0837 & 0.0291 & 0.0491 \\
 & Logistic & 0.0264 & 0.0123 & 0.0164 & 0.0213 & --- & 0.2138 & 0.0803 & 0.0283 & 0.0482 \\
 & PMM & 0.0265 & 0.0122 & 0.0161 & 0.0215 & --- & 0.2149 & 0.0810 & 0.0283 & 0.0485 \\
 & CART & 0.0267 & 0.0122 & 0.0164 & 0.0215 & --- & 0.2179 & 0.0820 & 0.0288 & 0.0494 \\
 & GERBIL & 0.0264 & 0.0122 & 0.0162 & 0.0214 & --- & 0.2130 & 0.0807 & 0.0283 & 0.0485 \\  \hline
\end{tabular}
}
\end{table}

\begin{table}[!ht]
\centering
{
\footnotesize
\renewcommand{\tabcolsep}{.12cm}
\caption{\footnotesize Coverage rates using six methods of imputation for the parameters of the fully-specified regression models of the simulated variables (where $\{\bm{X}_{12},\ldots,\bm{X}_{14}\}$ are binary indicators created from $\bm{X}_1$) across the 5,000 simulated datasets under a NMAR missingness mechanism. Rows indicate the outcome variable and columns indicate predictors.  
} \label{NMAR2}
\vspace{.12in}
\begin{tabular}{cccccccccccc}
\hline \hline
 & & Intercept & $\bm{X}_{12}$ & $\bm{X}_{13}$ & $\bm{X}_{14}$ & $\bm{X}_{2}$ & $\bm{X}_{3}$ & $\bm{X}_{4}$ & $\bm{X}_{5}$ & $\bm{X}_{6}$ \\  \hline
\multirow{8}{*}{sbgcop} & $\bm{X}_{12}$ & 0.9496 & --- & --- & --- & 0.9408 & 0.9458 & 0.9424 & 0.9364 & 0.9392 \\
 & $\bm{X}_{13}$ & 0.9492 & --- & --- & --- & 0.9534 & 0.9580 & 0.9434 & 0.9510 & 0.9562 \\
 & $\bm{X}_{14}$ & 0.9554 & --- & --- & --- & 0.9466 & 0.9552 & 0.9610 & 0.9414 & 0.9366 \\
 & $\bm{X}_{2}$ & 0.3628 & 0.9288 & 0.9534 & 0.9246 & --- & 0.9572 & 0.9222 & 0.7972 & 0.9234 \\
 & $\bm{X}_{3}$ & 0.0000 & 0.9324 & 0.9586 & 0.9394 & 0.8246 & --- & 0.8468 & 0.1098 & 0.9364 \\
 & $\bm{X}_{4}$ & 0.1456 & 0.9412 & 0.9402 & 0.9614 & 0.9452 & 0.8910 & --- & 0.6898 & 0.9286 \\
 & $\bm{X}_{5}$ & 0.7864 & 0.9278 & 0.9456 & 0.9336 & 0.8218 & 0.2978 & 0.6208 & --- & 0.9446 \\
 & $\bm{X}_{6}$ & 0.9622 & 0.9398 & 0.9564 & 0.9364 & 0.9056 & 0.9462 & 0.9302 & 0.9508 & --- \\  \hline
\multirow{8}{*}{jomo} & $\bm{X}_{12}$ & 0.9502 & --- & --- & --- & 0.9420 & 0.9502 & 0.9452 & 0.9560 & 0.9476 \\
 & $\bm{X}_{13}$ & 0.9360 & --- & --- & --- & 0.9540 & 0.9556 & 0.9470 & 0.9592 & 0.9562 \\
 & $\bm{X}_{14}$ & 0.9222 & --- & --- & --- & 0.9480 & 0.9456 & 0.9554 & 0.9496 & 0.9518 \\
 & $\bm{X}_{2}$ & 0.5550 & 0.9370 & 0.9544 & 0.9400 & --- & 0.9426 & 0.9260 & 0.7934 & 0.9470 \\
 & $\bm{X}_{3}$ & 0.0182 & 0.9380 & 0.9528 & 0.9392 & 0.8120 & --- & 0.9268 & 0.4702 & 0.9022 \\
 & $\bm{X}_{4}$ & 0.8874 & 0.9436 & 0.9476 & 0.9532 & 0.9410 & 0.9482 & --- & 0.8668 & 0.9470 \\
 & $\bm{X}_{5}$ & 0.9192 & 0.9468 & 0.9602 & 0.9554 & 0.7316 & 0.8872 & 0.9090 & --- & 0.9150 \\
 & $\bm{X}_{6}$ & 0.9480 & 0.9468 & 0.9576 & 0.9520 & 0.9464 & 0.9440 & 0.9472 & 0.8992 & --- \\  \hline
\multirow{8}{*}{Logistic} & $\bm{X}_{12}$ & 0.9426 & --- & --- & --- & 0.9400 & 0.9506 & 0.9466 & 0.9424 & 0.9420 \\
 & $\bm{X}_{13}$ & 0.9348 & --- & --- & --- & 0.9530 & 0.9510 & 0.9496 & 0.9512 & 0.9568 \\
 & $\bm{X}_{14}$ & 0.9372 & --- & --- & --- & 0.9490 & 0.9488 & 0.9512 & 0.9390 & 0.9478 \\
 & $\bm{X}_{2}$ & 0.7520 & 0.9362 & 0.9504 & 0.9374 & --- & 0.9506 & 0.9228 & 0.9114 & 0.9484 \\
 & $\bm{X}_{3}$ & 0.1498 & 0.9392 & 0.9524 & 0.9330 & 0.7400 & --- & 0.9236 & 0.8330 & 0.9148 \\
 & $\bm{X}_{4}$ & 0.5870 & 0.9460 & 0.9496 & 0.9520 & 0.9388 & 0.9516 & --- & 0.9502 & 0.9478 \\
 & $\bm{X}_{5}$ & 0.9478 & 0.9388 & 0.9492 & 0.9392 & 0.9290 & 0.9460 & 0.9470 & --- & 0.9484 \\
 & $\bm{X}_{6}$ & 0.9514 & 0.9432 & 0.9562 & 0.9470 & 0.9434 & 0.9478 & 0.9478 & 0.9500 & --- \\  \hline
\multirow{8}{*}{PMM} & $\bm{X}_{12}$ & 0.9614 & --- & --- & --- & 0.9622 & 0.9642 & 0.9656 & 0.9622 & 0.9452 \\
 & $\bm{X}_{13}$ & 0.9580 & --- & --- & --- & 0.9684 & 0.9684 & 0.9742 & 0.9670 & 0.9658 \\
 & $\bm{X}_{14}$ & 0.9276 & --- & --- & --- & 0.9446 & 0.9418 & 0.9474 & 0.9390 & 0.9356 \\
 & $\bm{X}_{2}$ & 0.7534 & 0.9542 & 0.9668 & 0.9378 & --- & 0.9386 & 0.9354 & 0.9014 & 0.9484 \\
 & $\bm{X}_{3}$ & 0.1422 & 0.9492 & 0.9664 & 0.9182 & 0.7582 & --- & 0.9088 & 0.8370 & 0.9080 \\
 & $\bm{X}_{4}$ & 0.4876 & 0.9672 & 0.9754 & 0.9466 & 0.9500 & 0.9510 & --- & 0.9330 & 0.9540 \\
 & $\bm{X}_{5}$ & 0.9518 & 0.9608 & 0.9672 & 0.9378 & 0.9316 & 0.9476 & 0.9312 & --- & 0.9520 \\
 & $\bm{X}_{6}$ & 0.9548 & 0.9416 & 0.9660 & 0.9360 & 0.9446 & 0.9454 & 0.9544 & 0.9536 & --- \\  \hline
\multirow{8}{*}{CART} & $\bm{X}_{12}$ & 0.9580 & --- & --- & --- & 0.9140 & 0.9518 & 0.9640 & 0.9634 & 0.9704 \\
 & $\bm{X}_{13}$ & 0.9560 & --- & --- & --- & 0.9550 & 0.9672 & 0.9858 & 0.9718 & 0.8844 \\
 & $\bm{X}_{14}$ & 0.9362 & --- & --- & --- & 0.9206 & 0.9506 & 0.9600 & 0.9526 & 0.9386 \\
 & $\bm{X}_{2}$ & 0.5746 & 0.8926 & 0.9502 & 0.9062 & --- & 0.9222 & 0.8948 & 0.8164 & 0.9076 \\
 & $\bm{X}_{3}$ & 0.0848 & 0.9096 & 0.9650 & 0.9008 & 0.6686 & --- & 0.8900 & 0.6744 & 0.8266 \\
 & $\bm{X}_{4}$ & 0.7432 & 0.9670 & 0.9858 & 0.9598 & 0.9098 & 0.8776 & --- & 0.8544 & 0.9136 \\
 & $\bm{X}_{5}$ & 0.9060 & 0.9578 & 0.9724 & 0.9532 & 0.8108 & 0.8992 & 0.8620 & --- & 0.8370 \\
 & $\bm{X}_{6}$ & 0.9576 & 0.9700 & 0.8856 & 0.9392 & 0.8968 & 0.9058 & 0.9106 & 0.8226 & --- \\  \hline
\multirow{8}{*}{GERBIL} & $\bm{X}_{12}$ & 0.9436 & --- & --- & --- & 0.9464 & 0.9488 & 0.9412 & 0.9486 & 0.9434 \\
 & $\bm{X}_{13}$ & 0.9404 & --- & --- & --- & 0.9536 & 0.9486 & 0.9486 & 0.9574 & 0.9560 \\
 & $\bm{X}_{14}$ & 0.9386 & --- & --- & --- & 0.9496 & 0.9470 & 0.9530 & 0.9372 & 0.9484 \\
 & $\bm{X}_{2}$ & 0.7730 & 0.9402 & 0.9544 & 0.9436 & --- & 0.9458 & 0.9238 & 0.9188 & 0.9468 \\
 & $\bm{X}_{3}$ & 0.1662 & 0.9440 & 0.9522 & 0.9338 & 0.7620 & --- & 0.9238 & 0.8436 & 0.9168 \\
 & $\bm{X}_{4}$ & 0.5688 & 0.9422 & 0.9468 & 0.9526 & 0.9402 & 0.9514 & --- & 0.9476 & 0.9460 \\
 & $\bm{X}_{5}$ & 0.9480 & 0.9456 & 0.9558 & 0.9416 & 0.9352 & 0.9488 & 0.9462 & --- & 0.9508 \\
 & $\bm{X}_{6}$ & 0.9494 & 0.9444 & 0.9568 & 0.9478 & 0.9436 & 0.9456 & 0.9476 & 0.9518 & --- \\  \hline
\end{tabular}
}
\end{table}

\begin{table}[!ht]
\centering
{
\footnotesize
\renewcommand{\tabcolsep}{.12cm}
\caption{\footnotesize rMSE using six methods of imputation for the parameters of the fully-specified regression models of the simulated variables (where $\{\bm{X}_{12},\ldots,\bm{X}_{14}\}$ are binary indicators created from $\bm{X}_1$) across the 5,000 simulated datasets under a NMAR missingness mechanism. Rows indicate the outcome variable and columns indicate predictors.  
} \label{NMAR3}
\vspace{.12in}
\begin{tabular}{cccccccccccc}
\hline \hline
 & & Intercept & $\bm{X}_{12}$ & $\bm{X}_{13}$ & $\bm{X}_{14}$ & $\bm{X}_{2}$ & $\bm{X}_{3}$ & $\bm{X}_{4}$ & $\bm{X}_{5}$ & $\bm{X}_{6}$ \\  \hline
\multirow{8}{*}{sbgcop} & $\bm{X}_{12}$ & 0.3275 & --- & --- & --- & 0.1031 & 0.1299 & 0.2530 & 0.1420 & 0.2123 \\
 & $\bm{X}_{13}$ & 0.3076 & --- & --- & --- & 0.0923 & 0.1205 & 0.2494 & 0.1300 & 0.1942 \\
 & $\bm{X}_{14}$ & 0.3306 & --- & --- & --- & 0.1047 & 0.1236 & 0.2315 & 0.1427 & 0.2199 \\
 & $\bm{X}_{2}$ & 0.2081 & 0.0692 & 0.0604 & 0.0710 & --- & 0.0279 & 0.0653 & 0.0507 & 0.0561 \\
 & $\bm{X}_{3}$ & 0.6318 & 0.0822 & 0.0712 & 0.0778 & 0.0384 & --- & 0.0971 & 0.1226 & 0.0621 \\
 & $\bm{X}_{4}$ & 1.1322 & 0.2578 & 0.2547 & 0.2347 & 0.1029 & 0.1497 & --- & 0.2409 & 0.2280 \\
 & $\bm{X}_{5}$ & 0.0686 & 0.0523 & 0.0474 & 0.0526 & 0.0270 & 0.0598 & 0.0876 & --- & 0.0401 \\
 & $\bm{X}_{6}$ & 0.2582 & 0.2118 & 0.1936 & 0.2200 & 0.0950 & 0.1037 & 0.2229 & 0.1116 & --- \\  \hline
\multirow{8}{*}{jomo} & $\bm{X}_{12}$ & 0.3343 & --- & --- & --- & 0.1032 & 0.1274 & 0.2468 & 0.1290 & 0.2037 \\
 & $\bm{X}_{13}$ & 0.3352 & --- & --- & --- & 0.0949 & 0.1247 & 0.2438 & 0.1215 & 0.1888 \\
 & $\bm{X}_{14}$ & 0.3849 & --- & --- & --- & 0.1044 & 0.1318 & 0.2438 & 0.1329 & 0.2056 \\
 & $\bm{X}_{2}$ & 0.1798 & 0.0666 & 0.0609 & 0.0669 & --- & 0.0295 & 0.0641 & 0.0501 & 0.0506 \\
 & $\bm{X}_{3}$ & 0.4300 & 0.0760 & 0.0714 & 0.0775 & 0.0394 & --- & 0.0728 & 0.0820 & 0.0673 \\
 & $\bm{X}_{4}$ & 0.4712 & 0.2493 & 0.2470 & 0.2458 & 0.1058 & 0.1263 & --- & 0.1689 & 0.2041 \\
 & $\bm{X}_{5}$ & 0.0498 & 0.0486 & 0.0445 & 0.0478 & 0.0318 & 0.0294 & 0.0548 & --- & 0.0435 \\
 & $\bm{X}_{6}$ & 0.2829 & 0.2032 & 0.1883 & 0.2054 & 0.0840 & 0.1055 & 0.2005 & 0.1249 & --- \\  \hline
\multirow{8}{*}{Logistic} & $\bm{X}_{12}$ & 0.3452 & --- & --- & --- & 0.1033 & 0.1294 & 0.2481 & 0.1391 & 0.2063 \\
 & $\bm{X}_{13}$ & 0.3379 & --- & --- & --- & 0.0948 & 0.1288 & 0.2453 & 0.1297 & 0.1906 \\
 & $\bm{X}_{14}$ & 0.3659 & --- & --- & --- & 0.1035 & 0.1339 & 0.2467 & 0.1437 & 0.2093 \\
 & $\bm{X}_{2}$ & 0.1417 & 0.0670 & 0.0611 & 0.0666 & --- & 0.0295 & 0.0657 & 0.0396 & 0.0507 \\
 & $\bm{X}_{3}$ & 0.3389 & 0.0763 & 0.0724 & 0.0797 & 0.0446 & --- & 0.0767 & 0.0536 & 0.0653 \\
 & $\bm{X}_{4}$ & 0.7697 & 0.2505 & 0.2490 & 0.2488 & 0.1076 & 0.1271 & --- & 0.1364 & 0.2085 \\
 & $\bm{X}_{5}$ & 0.0434 & 0.0493 & 0.0457 & 0.0505 & 0.0204 & 0.0233 & 0.0465 & --- & 0.0389 \\
 & $\bm{X}_{6}$ & 0.2899 & 0.2063 & 0.1906 & 0.2093 & 0.0840 & 0.1070 & 0.2061 & 0.1114 & --- \\  \hline
\multirow{8}{*}{PMM} & $\bm{X}_{12}$ & 0.3255 & --- & --- & --- & 0.0944 & 0.1210 & 0.2210 & 0.1285 & 0.2097 \\
 & $\bm{X}_{13}$ & 0.3048 & --- & --- & --- & 0.0865 & 0.1172 & 0.2072 & 0.1193 & 0.1791 \\
 & $\bm{X}_{14}$ & 0.3840 & --- & --- & --- & 0.1046 & 0.1384 & 0.2530 & 0.1462 & 0.2243 \\
 & $\bm{X}_{2}$ & 0.1416 & 0.0620 & 0.0560 & 0.0676 & --- & 0.0303 & 0.0623 & 0.0402 & 0.0503 \\
 & $\bm{X}_{3}$ & 0.3399 & 0.0743 & 0.0651 & 0.0839 & 0.0434 & --- & 0.0792 & 0.0529 & 0.0674 \\
 & $\bm{X}_{4}$ & 0.8488 & 0.2224 & 0.2093 & 0.2545 & 0.1015 & 0.1250 & --- & 0.1486 & 0.2010 \\
 & $\bm{X}_{5}$ & 0.0436 & 0.0454 & 0.0420 & 0.0511 & 0.0201 & 0.0233 & 0.0495 & --- & 0.0384 \\
 & $\bm{X}_{6}$ & 0.2888 & 0.2112 & 0.1786 & 0.2246 & 0.0835 & 0.1063 & 0.1992 & 0.1105 & --- \\  \hline
\multirow{8}{*}{CART} & $\bm{X}_{12}$ & 0.3185 & --- & --- & --- & 0.1125 & 0.1194 & 0.2020 & 0.1219 & 0.1657 \\
 & $\bm{X}_{13}$ & 0.3077 & --- & --- & --- & 0.0910 & 0.1062 & 0.1682 & 0.1117 & 0.2220 \\
 & $\bm{X}_{14}$ & 0.3619 & --- & --- & --- & 0.1112 & 0.1227 & 0.2117 & 0.1296 & 0.2028 \\
 & $\bm{X}_{2}$ & 0.1708 & 0.0744 & 0.0588 & 0.0723 & --- & 0.0308 & 0.0660 & 0.0471 & 0.0552 \\
 & $\bm{X}_{3}$ & 0.3699 & 0.0792 & 0.0607 & 0.0804 & 0.0474 & --- & 0.0753 & 0.0652 & 0.0754 \\
 & $\bm{X}_{4}$ & 0.5919 & 0.2026 & 0.1693 & 0.2134 & 0.1094 & 0.1475 & --- & 0.1659 & 0.2022 \\
 & $\bm{X}_{5}$ & 0.0488 & 0.0442 & 0.0398 & 0.0465 & 0.0270 & 0.0265 & 0.0578 & --- & 0.0497 \\
 & $\bm{X}_{6}$ & 0.2575 & 0.1648 & 0.2209 & 0.2020 & 0.0912 & 0.1117 & 0.1995 & 0.1415 & --- \\  \hline
\multirow{8}{*}{GERBIL} & $\bm{X}_{12}$ & 0.3457 & --- & --- & --- & 0.1031 & 0.1304 & 0.2510 & 0.1388 & 0.2083 \\
 & $\bm{X}_{13}$ & 0.3353 & --- & --- & --- & 0.0945 & 0.1274 & 0.2475 & 0.1279 & 0.1905 \\
 & $\bm{X}_{14}$ & 0.3673 & --- & --- & --- & 0.1032 & 0.1362 & 0.2461 & 0.1432 & 0.2108 \\
 & $\bm{X}_{2}$ & 0.1370 & 0.0662 & 0.0603 & 0.0658 & --- & 0.0298 & 0.0654 & 0.0387 & 0.0505 \\
 & $\bm{X}_{3}$ & 0.3264 & 0.0755 & 0.0706 & 0.0788 & 0.0433 & --- & 0.0764 & 0.0512 & 0.0648 \\
 & $\bm{X}_{4}$ & 0.7816 & 0.2554 & 0.2531 & 0.2496 & 0.1084 & 0.1293 & --- & 0.1378 & 0.2132 \\
 & $\bm{X}_{5}$ & 0.0435 & 0.0489 & 0.0448 & 0.0501 & 0.0200 & 0.0228 & 0.0464 & --- & 0.0388 \\
 & $\bm{X}_{6}$ & 0.2951 & 0.2074 & 0.1896 & 0.2105 & 0.0837 & 0.1080 & 0.2083 & 0.1114 & --- \\  \hline
\end{tabular}
}
\end{table}

\begin{table}[!ht]
\centering
{
\scriptsize
\renewcommand{\tabcolsep}{.12cm}
\caption{\footnotesize rMSE using six methods of imputation for elements of the variance-covariance matrix calculated with the simulated variables (where $\{\bm{X}_{11},\ldots,\bm{X}_{14}\}$ are binary indicators created from $\bm{X}_1$) across the 5,000 simulated datasets under a NMAR missingness mechanism. 
} \label{NMAR4}
\vspace{.12in}
\begin{tabular}{ccccccccccccc}
\hline \hline
 & & $\bm{X}_{11}$ & $\bm{X}_{12}$ & $\bm{X}_{13}$ & $\bm{X}_{14}$ & $\bm{X}_{2}$ & $\bm{X}_{3}$ & $\bm{X}_{4}$ & $\bm{X}_{5}$ & $\bm{X}_{6}$ \\  \hline
\multirow{9}{*}{sbgcop} & $\bm{X}_{11}$ & 0.0155 & --- & --- & --- & --- & --- & --- & --- & --- \\
 & $\bm{X}_{12}$ & 0.0073 & 0.0061 & --- & --- & --- & --- & --- & --- & --- \\
 & $\bm{X}_{13}$ & 0.0050 & 0.0047 & 0.0086 & --- & --- & --- & --- & --- & --- \\
 & $\bm{X}_{14}$ & 0.0049 & 0.0049 & 0.0076 & 0.0080 & --- & --- & --- & --- & --- \\
 & $\bm{X}_{2}$ & 0.0148 & 0.0121 & 0.0123 & 0.0110 & --- & --- & --- & --- & --- \\
 & $\bm{X}_{3}$ & 0.0325 & 0.0214 & 0.0138 & 0.0225 & 0.1128 & 0.2052 & --- & --- & --- \\
 & $\bm{X}_{4}$ & 0.0082 & 0.0075 & 0.0072 & 0.0063 & 0.0142 & 0.0522 & 0.0056 & --- & --- \\
 & $\bm{X}_{5}$ & 0.0110 & 0.0088 & 0.0081 & 0.0080 & 0.0222 & 0.1054 & 0.0099 & 0.0199 & --- \\
 & $\bm{X}_{6}$ & 0.0065 & 0.0072 & 0.0067 & 0.0067 & 0.0154 & 0.0171 & 0.0096 & 0.0130 & 0.0017 \\  \hline
\multirow{9}{*}{jomo} & $\bm{X}_{11}$ & 0.0140 & --- & --- & --- & --- & --- & --- & --- & --- \\
 & $\bm{X}_{12}$ & 0.0080 & 0.0062 & --- & --- & --- & --- & --- & --- & --- \\
 & $\bm{X}_{13}$ & 0.0044 & 0.0035 & 0.0079 & --- & --- & --- & --- & --- & --- \\
 & $\bm{X}_{14}$ & 0.0037 & 0.0042 & 0.0085 & 0.0101 & --- & --- & --- & --- & --- \\
 & $\bm{X}_{2}$ & 0.0146 & 0.0119 & 0.0133 & 0.0115 & --- & --- & --- & --- & --- \\
 & $\bm{X}_{3}$ & 0.0278 & 0.0176 & 0.0120 & 0.0172 & 0.0860 & 0.1848 & --- & --- & --- \\
 & $\bm{X}_{4}$ & 0.0090 & 0.0078 & 0.0068 & 0.0069 & 0.0176 & 0.0429 & 0.0074 & --- & --- \\
 & $\bm{X}_{5}$ & 0.0121 & 0.0087 & 0.0083 & 0.0082 & 0.0220 & 0.0770 & 0.0210 & 0.0198 & --- \\
 & $\bm{X}_{6}$ & 0.0065 & 0.0067 & 0.0066 & 0.0066 & 0.0152 & 0.0177 & 0.0090 & 0.0106 & 0.0027 \\  \hline
\multirow{9}{*}{Logistic} & $\bm{X}_{11}$ & 0.0140 & --- & --- & --- & --- & --- & --- & --- & --- \\
 & $\bm{X}_{12}$ & 0.0080 & 0.0063 & --- & --- & --- & --- & --- & --- & --- \\
 & $\bm{X}_{13}$ & 0.0044 & 0.0034 & 0.0079 & --- & --- & --- & --- & --- & --- \\
 & $\bm{X}_{14}$ & 0.0036 & 0.0042 & 0.0085 & 0.0102 & --- & --- & --- & --- & --- \\
 & $\bm{X}_{2}$ & 0.0139 & 0.0120 & 0.0136 & 0.0121 & --- & --- & --- & --- & --- \\
 & $\bm{X}_{3}$ & 0.0250 & 0.0170 & 0.0122 & 0.0150 & 0.0788 & 0.1734 & --- & --- & --- \\
 & $\bm{X}_{4}$ & 0.0080 & 0.0078 & 0.0072 & 0.0070 & 0.0146 & 0.0355 & 0.0068 & --- & --- \\
 & $\bm{X}_{5}$ & 0.0102 & 0.0089 & 0.0091 & 0.0084 & 0.0207 & 0.0567 & 0.0117 & 0.0196 & --- \\
 & $\bm{X}_{6}$ & 0.0065 & 0.0067 & 0.0066 & 0.0067 & 0.0153 & 0.0178 & 0.0090 & 0.0109 & 0.0026 \\  \hline
\multirow{9}{*}{PMM} & $\bm{X}_{11}$ & 0.0140 & --- & --- & --- & --- & --- & --- & --- & --- \\
 & $\bm{X}_{12}$ & 0.0080 & 0.0062 & --- & --- & --- & --- & --- & --- & --- \\
 & $\bm{X}_{13}$ & 0.0045 & 0.0034 & 0.0078 & --- & --- & --- & --- & --- & --- \\
 & $\bm{X}_{14}$ & 0.0036 & 0.0043 & 0.0085 & 0.0102 & --- & --- & --- & --- & --- \\
 & $\bm{X}_{2}$ & 0.0142 & 0.0129 & 0.0114 & 0.0116 & --- & --- & --- & --- & --- \\
 & $\bm{X}_{3}$ & 0.0257 & 0.0182 & 0.0114 & 0.0160 & 0.0808 & 0.1797 & --- & --- & --- \\
 & $\bm{X}_{4}$ & 0.0081 & 0.0080 & 0.0065 & 0.0072 & 0.0152 & 0.0376 & 0.0069 & --- & --- \\
 & $\bm{X}_{5}$ & 0.0104 & 0.0093 & 0.0078 & 0.0082 & 0.0206 & 0.0576 & 0.0110 & 0.0197 & --- \\
 & $\bm{X}_{6}$ & 0.0063 & 0.0063 & 0.0071 & 0.0066 & 0.0152 & 0.0178 & 0.0088 & 0.0109 & 0.0027 \\  \hline
\multirow{9}{*}{CART} & $\bm{X}_{11}$ & 0.0141 & --- & --- & --- & --- & --- & --- & --- & --- \\
 & $\bm{X}_{12}$ & 0.0081 & 0.0063 & --- & --- & --- & --- & --- & --- & --- \\
 & $\bm{X}_{13}$ & 0.0045 & 0.0035 & 0.0079 & --- & --- & --- & --- & --- & --- \\
 & $\bm{X}_{14}$ & 0.0037 & 0.0043 & 0.0086 & 0.0102 & --- & --- & --- & --- & --- \\
 & $\bm{X}_{2}$ & 0.0140 & 0.0121 & 0.0136 & 0.0121 & --- & --- & --- & --- & --- \\
 & $\bm{X}_{3}$ & 0.0306 & 0.0195 & 0.0116 & 0.0175 & 0.0820 & 0.1804 & --- & --- & --- \\
 & $\bm{X}_{4}$ & 0.0105 & 0.0084 & 0.0059 & 0.0064 & 0.0152 & 0.0372 & 0.0071 & --- & --- \\
 & $\bm{X}_{5}$ & 0.0120 & 0.0094 & 0.0085 & 0.0080 & 0.0209 & 0.0635 & 0.0165 & 0.0199 & --- \\
 & $\bm{X}_{6}$ & 0.0080 & 0.0061 & 0.0076 & 0.0069 & 0.0156 & 0.0204 & 0.0097 & 0.0109 & 0.0028 \\  \hline
\multirow{9}{*}{GERBIL} & $\bm{X}_{11}$ & 0.0140 & --- & --- & --- & --- & --- & --- & --- & --- \\
 & $\bm{X}_{12}$ & 0.0080 & 0.0063 & --- & --- & --- & --- & --- & --- & --- \\
 & $\bm{X}_{13}$ & 0.0045 & 0.0034 & 0.0078 & --- & --- & --- & --- & --- & --- \\
 & $\bm{X}_{14}$ & 0.0036 & 0.0042 & 0.0085 & 0.0102 & --- & --- & --- & --- & --- \\
 & $\bm{X}_{2}$ & 0.0137 & 0.0122 & 0.0129 & 0.0119 & --- & --- & --- & --- & --- \\
 & $\bm{X}_{3}$ & 0.0234 & 0.0176 & 0.0118 & 0.0150 & 0.0762 & 0.1707 & --- & --- & --- \\
 & $\bm{X}_{4}$ & 0.0080 & 0.0078 & 0.0072 & 0.0068 & 0.0145 & 0.0346 & 0.0069 & --- & --- \\
 & $\bm{X}_{5}$ & 0.0100 & 0.0090 & 0.0086 & 0.0082 & 0.0206 & 0.0545 & 0.0115 & 0.0196 & --- \\
 & $\bm{X}_{6}$ & 0.0065 & 0.0068 & 0.0065 & 0.0067 & 0.0153 & 0.0176 & 0.0090 & 0.0110 & 0.0027 \\  \hline
\end{tabular}
}
\end{table}

\begin{table}[!ht]
\centering
{
\footnotesize
\renewcommand{\tabcolsep}{.12cm}
\caption{\footnotesize rMSE using six methods of imputation for the squared standard errors of the parameters of the fully-specified regression models with the simulated variables (where $\{\bm{X}_{12},\ldots,\bm{X}_{14}\}$ are binary indicators created from $\bm{X}_1$) across the 5,000 simulated datasets under a NMAR missingness mechanism. Rows indicate the outcome variable and columns indicate predictors.  
} \label{NMAR5}
\vspace{.12in}
\begin{tabular}{cccccccccccc}
\hline \hline
 & & Intercept & $\bm{X}_{12}$ & $\bm{X}_{13}$ & $\bm{X}_{14}$ & $\bm{X}_{2}$ & $\bm{X}_{3}$ & $\bm{X}_{4}$ & $\bm{X}_{5}$ & $\bm{X}_{6}$ \\  \hline
\multirow{8}{*}{sbgcop} & $\bm{X}_{12}$ & 0.2070 & --- & --- & --- & 0.0773 & 0.0768 & 0.1276 & 0.0993 & 0.1192 \\
 & $\bm{X}_{13}$ & 0.2007 & --- & --- & --- & 0.0744 & 0.0738 & 0.1257 & 0.0962 & 0.1151 \\
 & $\bm{X}_{14}$ & 0.2154 & --- & --- & --- & 0.0804 & 0.0797 & 0.1330 & 0.1031 & 0.1230 \\
 & $\bm{X}_{2}$ & 0.0708 & 0.0500 & 0.0483 & 0.0517 & --- & 0.0206 & 0.0400 & 0.0283 & 0.0352 \\
 & $\bm{X}_{3}$ & 0.0722 & 0.0501 & 0.0485 & 0.0520 & 0.0206 & --- & 0.0402 & 0.0280 & 0.0351 \\
 & $\bm{X}_{4}$ & 0.1807 & 0.1280 & 0.1262 & 0.1336 & 0.0647 & 0.0642 & --- & 0.0822 & 0.0995 \\
 & $\bm{X}_{5}$ & 0.0339 & 0.0380 & 0.0367 & 0.0394 & 0.0160 & 0.0159 & 0.0305 & --- & 0.0270 \\
 & $\bm{X}_{6}$ & 0.1669 & 0.1190 & 0.1150 & 0.1229 & 0.0561 & 0.0561 & 0.0987 & 0.0725 & --- \\  \hline
\multirow{8}{*}{jomo} & $\bm{X}_{12}$ & 0.2042 & --- & --- & --- & 0.0772 & 0.0764 & 0.1278 & 0.0994 & 0.1192 \\
 & $\bm{X}_{13}$ & 0.1988 & --- & --- & --- & 0.0743 & 0.0736 & 0.1261 & 0.0964 & 0.1153 \\
 & $\bm{X}_{14}$ & 0.2149 & --- & --- & --- & 0.0804 & 0.0795 & 0.1339 & 0.1036 & 0.1235 \\
 & $\bm{X}_{2}$ & 0.0707 & 0.0500 & 0.0484 & 0.0518 & --- & 0.0206 & 0.0400 & 0.0283 & 0.0352 \\
 & $\bm{X}_{3}$ & 0.0724 & 0.0502 & 0.0486 & 0.0521 & 0.0206 & --- & 0.0403 & 0.0280 & 0.0352 \\
 & $\bm{X}_{4}$ & 0.1849 & 0.1284 & 0.1268 & 0.1346 & 0.0648 & 0.0640 & --- & 0.0827 & 0.0999 \\
 & $\bm{X}_{5}$ & 0.0340 & 0.0380 & 0.0367 & 0.0394 & 0.0160 & 0.0159 & 0.0304 & --- & 0.0269 \\
 & $\bm{X}_{6}$ & 0.1660 & 0.1190 & 0.1151 & 0.1234 & 0.0561 & 0.0560 & 0.0991 & 0.0727 & --- \\  \hline
\multirow{8}{*}{Logistic} & $\bm{X}_{12}$ & 0.2027 & --- & --- & --- & 0.0772 & 0.0763 & 0.1272 & 0.0988 & 0.1191 \\
 & $\bm{X}_{13}$ & 0.1975 & --- & --- & --- & 0.0743 & 0.0734 & 0.1256 & 0.0959 & 0.1153 \\
 & $\bm{X}_{14}$ & 0.2124 & --- & --- & --- & 0.0804 & 0.0793 & 0.1332 & 0.1029 & 0.1233 \\
 & $\bm{X}_{2}$ & 0.0705 & 0.0500 & 0.0484 & 0.0518 & --- & 0.0206 & 0.0400 & 0.0283 & 0.0352 \\
 & $\bm{X}_{3}$ & 0.0725 & 0.0502 & 0.0486 & 0.0522 & 0.0206 & --- & 0.0403 & 0.0280 & 0.0352 \\
 & $\bm{X}_{4}$ & 0.1807 & 0.1277 & 0.1260 & 0.1337 & 0.0647 & 0.0638 & --- & 0.0821 & 0.0995 \\
 & $\bm{X}_{5}$ & 0.0340 & 0.0380 & 0.0367 & 0.0395 & 0.0160 & 0.0159 & 0.0305 & --- & 0.0270 \\
 & $\bm{X}_{6}$ & 0.1652 & 0.1189 & 0.1151 & 0.1232 & 0.0561 & 0.0559 & 0.0987 & 0.0724 & --- \\  \hline
\multirow{8}{*}{PMM} & $\bm{X}_{12}$ & 0.2034 & --- & --- & --- & 0.0772 & 0.0763 & 0.1274 & 0.0988 & 0.1192 \\
 & $\bm{X}_{13}$ & 0.1975 & --- & --- & --- & 0.0743 & 0.0734 & 0.1257 & 0.0959 & 0.1154 \\
 & $\bm{X}_{14}$ & 0.2128 & --- & --- & --- & 0.0804 & 0.0793 & 0.1333 & 0.1028 & 0.1233 \\
 & $\bm{X}_{2}$ & 0.0705 & 0.0500 & 0.0484 & 0.0518 & --- & 0.0206 & 0.0400 & 0.0283 & 0.0352 \\
 & $\bm{X}_{3}$ & 0.0725 & 0.0502 & 0.0486 & 0.0522 & 0.0206 & --- & 0.0403 & 0.0280 & 0.0352 \\
 & $\bm{X}_{4}$ & 0.1803 & 0.1278 & 0.1262 & 0.1339 & 0.0647 & 0.0639 & --- & 0.0821 & 0.0995 \\
 & $\bm{X}_{5}$ & 0.0340 & 0.0380 & 0.0367 & 0.0395 & 0.0160 & 0.0159 & 0.0305 & --- & 0.0270 \\
 & $\bm{X}_{6}$ & 0.1653 & 0.1190 & 0.1152 & 0.1232 & 0.0561 & 0.0559 & 0.0988 & 0.0724 & --- \\  \hline
\multirow{8}{*}{CART} & $\bm{X}_{12}$ & 0.2035 & --- & --- & --- & 0.0772 & 0.0764 & 0.1280 & 0.0992 & 0.1196 \\
 & $\bm{X}_{13}$ & 0.1978 & --- & --- & --- & 0.0743 & 0.0735 & 0.1265 & 0.0962 & 0.1157 \\
 & $\bm{X}_{14}$ & 0.2137 & --- & --- & --- & 0.0805 & 0.0795 & 0.1341 & 0.1033 & 0.1239 \\
 & $\bm{X}_{2}$ & 0.0706 & 0.0500 & 0.0484 & 0.0518 & --- & 0.0206 & 0.0400 & 0.0283 & 0.0353 \\
 & $\bm{X}_{3}$ & 0.0724 & 0.0502 & 0.0486 & 0.0522 & 0.0206 & --- & 0.0403 & 0.0280 & 0.0352 \\
 & $\bm{X}_{4}$ & 0.1832 & 0.1285 & 0.1271 & 0.1347 & 0.0647 & 0.0639 & --- & 0.0824 & 0.0999 \\
 & $\bm{X}_{5}$ & 0.0340 & 0.0380 & 0.0367 & 0.0395 & 0.0160 & 0.0159 & 0.0305 & --- & 0.0270 \\
 & $\bm{X}_{6}$ & 0.1659 & 0.1194 & 0.1154 & 0.1238 & 0.0561 & 0.0560 & 0.0992 & 0.0726 & --- \\  \hline
\multirow{8}{*}{GERBIL} & $\bm{X}_{12}$ & 0.2026 & --- & --- & --- & 0.0772 & 0.0762 & 0.1272 & 0.0988 & 0.1191 \\
 & $\bm{X}_{13}$ & 0.1974 & --- & --- & --- & 0.0743 & 0.0734 & 0.1255 & 0.0959 & 0.1152 \\
 & $\bm{X}_{14}$ & 0.2124 & --- & --- & --- & 0.0804 & 0.0793 & 0.1331 & 0.1028 & 0.1232 \\
 & $\bm{X}_{2}$ & 0.0705 & 0.0500 & 0.0484 & 0.0518 & --- & 0.0206 & 0.0400 & 0.0283 & 0.0352 \\
 & $\bm{X}_{3}$ & 0.0725 & 0.0502 & 0.0486 & 0.0522 & 0.0206 & --- & 0.0403 & 0.0280 & 0.0352 \\
 & $\bm{X}_{4}$ & 0.1802 & 0.1275 & 0.1259 & 0.1337 & 0.0646 & 0.0638 & --- & 0.0821 & 0.0994 \\
 & $\bm{X}_{5}$ & 0.0340 & 0.0380 & 0.0367 & 0.0395 & 0.0160 & 0.0159 & 0.0305 & --- & 0.0270 \\
 & $\bm{X}_{6}$ & 0.1651 & 0.1189 & 0.1150 & 0.1232 & 0.0561 & 0.0559 & 0.0987 & 0.0724 & --- \\  \hline
\end{tabular}
}
\end{table}

\end{singlespace}

\end{document}